\documentclass[acmsmall,screen,nonacm]{acmart}

\usepackage{booktabs} 
\usepackage{subcaption}
\usepackage{mathtools}
\usepackage{multicol}
\usepackage{color,soul}
\sethlcolor{white}
\usepackage{makecell}
\usepackage{colortbl}
\usepackage{tabularx,colortbl}
\usepackage{caption}
\usepackage{booktabs}
\usepackage{multirow}
\usepackage{tikz}
\usepackage{multicol}
\usepackage{enumitem}
\usepackage{url}
\usepackage{graphics}
\usepackage{float}

\renewcommand\footnotetextcopyrightpermission[1]{}

\begin{document}

\title{Why and When: Understanding System Initiative during Conversational Collaborative Search}

\author{Sandeep Avula}
\authornote{Work done while at the University of North Carolina at Chapel Hill}
\email{sandeavu@amazon.com}
\affiliation{
\institution{Amazon}
\country{USA}
}

\author{Bogeum Choi}
\email{bochoi@unc.edu}
\affiliation{
\institution{University of North Carolina at Chapel Hill}
\country{USA}
}

\author{Jaime Arguello}
\email{jarguello@unc.edu}
\affiliation{
\institution{University of North Carolina at Chapel Hill}
\country{USA}
}

\begin{abstract}
In the last decade, conversational search has attracted considerable attention. However, most research has focused on systems that can support a \emph{single} searcher. In this paper, we explore how systems can support \emph{multiple} searchers collaborating over an instant messaging platform (i.e., Slack). We present a ``Wizard of Oz'' study in which 27 participant pairs collaborated on three information-seeking tasks over Slack. Participants were unable to search on their own and had to gather information by interacting with a \emph{searchbot} directly from the Slack channel. The role of the searchbot was played by a reference librarian.  Conversational search systems must be capable of engaging in \emph{mixed-initiative} interaction by taking and relinquishing control of the conversation to fulfill different objectives. Discourse analysis research suggests that conversational agents can take \emph{two} levels of initiative: dialog- and task-level initiative. Agents take dialog-level initiative to establish mutual belief between agents and task-level initiative to influence the goals of the other agents. During the study, participants were exposed to three experimental conditions in which the searchbot could take different levels of initiative: (1) no initiative, (2) only dialog-level initiative, and (3) both dialog- and task-level initiative. In this paper, we focus on understanding the Wizard's actions. Specifically, we focus on understanding the Wizard's motivations for taking initiative and their rationale for the timing of each intervention. To gain insights about the Wizard's actions, we conducted a stimulated recall interview with the Wizard. We present findings from a qualitative analysis of this interview data and discuss implications for designing conversational search systems to support collaborative search. 
\end{abstract}


\maketitle

\section{Introduction}\label{sec:intro}

Our work in this paper lies at the intersection of two areas of ongoing research: collaborative search and conversational search.  The goal of \emph{collaborative search} is to develop systems to support \emph{multiple} collaborators working together on a task that involves information seeking.  To date, the most common approach has been to develop \emph{standalone} systems that include a search interface and peripheral tools for collaborators to communicate, share information, and gain awareness of each other's activities.  Prior research has found that such standalone systems can provide important benefits, such as helping collaborators delegate tasks~\cite{Paul2009}, review each other's work~\cite{Htun2017,Capra2012}, avoid duplicating effort~\cite{Morris2007,Putra2018,Shah2013,Capra2012}, and track their progress~\cite{Shah2013}. However, despite their benefits, standalone systems have not been widely adopted~\cite{Morris2013,Hearst2014}. In a survey by Morris~\cite{Morris2013}, about half of the respondents reported doing collaborative searches at least once a week. However, all of the respondents reported using \emph{non-integrated} tools (e.g., web search engines and communication tools such as email and instant messaging) to conduct their collaborative searches.  Based on these findings, Morris proposed that future work should focus on embedding search technology into communication platforms that people already use to collaborate.  In this paper, we study how a conversational search system might support collaborations over an instant messaging platform (i.e., Slack).

In the last decade, conversational search has attracted considerable attention. The goal of \emph{conversational search} is to develop systems that users can interact with using dialog. Most research has focused on understanding how systems can support a \emph{single} searcher~\cite{vtyurina2017exploring,trippas2017people,budzianowski2018multiwoz,aliannejadi2019asking,vakulenko2021large}. Our research investigates conversational search in the context of a multiparty collaboration. 

Research in conversational search begs the question: What makes a search system \emph{conversational}? Radlinski and Craswell~\cite{Radlinski2017} enumerated five capabilities necessary for a search system to be considered ``conversational''.  One of these is the ability for the system to engage in \emph{mixed-initiative} interaction.  In other words, to be conversational, a search system must be capable of taking control of a conversation to fulfill different objectives.  Discourse analysis research has studied mixed-initiative interactions for many years~\cite{Walker1990}. Chu-Carroll and Brown~\cite{chu1997tracking} argued that task-oriented dialogs involve two levels of initiative: dialog- and task-level initiative.  Agents takes \emph{dialog-level} initiative to establish mutual belief between agents.  For example, taking dialog-level initiative might involve asking a clarification question in response to a request or recommendation.  Conversely, agents take \emph{task-level} initiative to influence the goals of the other agents.  For example, taking task-level initiative might involve proposing a new approach to the task. In this research, we investigate \emph{why} and \emph{when} a search system should take dialog- and task-level initiative to support collaborators during an information-seeking task.

We present a ``Wizard of Oz'' study in which 27 pairs of participants completed three collaborative search tasks over a messaging platform (i.e. Slack).  All three \emph{decision-making} tasks required pairs of participants to consider different alternatives and make a joint selection.  To gather information, participants had to interact with a so-called \emph{searchbot} directly from Slack.  The role of the searchbot was played by a reference librarian (referred to as the ``Wizard'').  The study used a within-subjects design and involved \emph{three} experimental conditions in which the Wizard could take \emph{different} levels of initiative during the task.  In all three conditions, participants had to gather information by issuing \emph{search requests} to the Wizard directly from the Slack channel.  In response to each search request, the Wizard searched the web and embedded a single search result directly in the Slack channel. In the \textsc{BotInfo} condition, the Wizard could not take any level of initiative.  In the \textsc{BotDialog} condition, the Wizard could only take dialog-level initiative by asking one or more clarification questions in response to a search request.  Finally, in the \textsc{BotTask} condition, the Wizard could take both dialog- and task-level initiative. That is, in addition to asking clarification questions, the Wizard in the \textsc{BotTask} condition could take initiative by providing task-level suggestions.  The Wizard could provide task-level suggestions either in response to a search request or by \emph{proactively} intervening in the conversation without being asked.

Our research in this paper builds on our own prior work. In a previous paper~\cite{Avula2021}, we reported on results from the same study.  However, in that paper, we focused on the effects of the searchbot condition on participants' perceptions and behaviors. Our results found benefits and challenges from systems that can take both dialog- and task-level initiative during a collaborative search session.  Specifically, in the \textsc{BotTask} condition, participants perceived the task to be more challenging but also explored a broader range of dimensions while evaluating alternatives.  In other words, task-level suggestions made the task more difficult but also influenced participants to consider new ideas, which can be viewed as a positive outcome. In the current paper, we present a completely novel qualitative analysis of the Wizard's responses during a stimulated-recall interview conducted at the end of each study session.  During this interview, the study moderator revisited \emph{every} instance in which the Wizard took the initiative in the \textsc{BotDialog} and \textsc{BotTask} conditions.  For each instance, the Wizard was asked questions about: (1) their \emph{motivation} for taking the initiative---What were they hoping to achieve and why?---and (2) their rationale for the \emph{timing} of the intervention---Why did they think that the participants would perceive the intervention as appropriate and not disruptive?  The Wizard's responses to these questions were analyzed using qualitative techniques to address two main research questions:

\begin{itemize}[leftmargin=*]
\item \textbf{RQ1 (Motivations):} What were the different motivations that prompted the Wizard to intervene with a clarification question or a task-level suggestion? In other words, when the Wizard intervened, what were they hoping to achieve and why?
\item \textbf{RQ2 (Timing):} How did the Wizard justify the timing of their interventions? In other words, why and how did the Wizard decide that the point of intervention would be perceived by participants as being appropriate versus disruptive?
\end{itemize}

Developing a fully conversational search system is a daunting endeavor.  A fundamental question is: What should the system be capable of doing?  To address this question, prior observational studies have aimed at understanding how a \emph{human} agent plays the role of a conversation search system~\cite{trippas2017people,trippas2020towards,vtyurina2017exploring,vtyurina2018exploring}.  However, to our knowledge, prior studies has focused on systems that can support a \emph{single} searcher.  In this work, we observe how a human agent can support \emph{multiple} searchers working together.  In this respect, our results provide insights into what a conversational collaborative search system should be capable of doing, how, and why. Additionally, understanding \textit{why} and \textit{when} a conversational agent should take initiative during collaborations becomes even more important in the age of Large Language Models (LLMs). Integrating LLMs into chatbots on platforms such as Slack, Discord, and Microsoft Teams that support collaborations between large groups of users is a natural application of their abilities. By conducting research in this area, we gain insights into how proactive agents can be used to support and improve group collaborations.

Our results found several important trends associated with: (RQ1) reasons for \emph{why} a system might decide to take initiative and (RQ2) factors that a system might consider in deciding whether the timing of an intervention is appropriate.

\textbf{Motivations:} In terms of clarification questions, our results suggest that a system might need to take dialog-level initiative in order to: (1) learn about the collaborators' problem space (e.g., constraints that might influence relevance criteria); (2) ask about ambiguous or subjective terminology in a search request; (3) ask about preferences/constraints mentioned in the conversation but not explicitly included in a search request; (4) ask about challenging aspects of a search request that might need to be modified; (5) ask about facet-values that might be useful for reducing the search space; and (6) repair a previous clarification attempt that was misunderstood by the collaborators.

In terms of task-based suggestions, our results suggest that a system might need to intervene in order to: (1) propose \emph{new} alternatives that seem relevant based on the conversation; (2) explain why or how a returned search result is relevant; (3) correct misconceptions that collaborators might have based on their conversation; (4) help collaborators decompose the task into more manageable subtasks; and (5) ask whether assistance is needed in situations where the collaborators seem idle or stuck.

\textbf{Timing:}  Our results suggest that a system might need to consider different factors in deciding whether the timing of an intervention will be perceived as appropriate versus disruptive.  Specifically, these factors can be grouped into three categories.  First, some factors relate to a sense of \emph{urgency}.  In such cases, from the Wizard's perspective, the benefit of the intervention outweighed the risk of the intervention being perceived as disruptive.  Examples included decisions to intervene in order to: (1) clarify a problematic search request; (2) provide an update about work-in-progress; (3) clarify an important misconception; (4) encourage collaborators to change strategies (or pursue a newly adopted strategy) after a period of struggle; and (5) inquire about a potential change to the current search request based on the things mentioned in the ongoing conversation.  Second, some factors relate to the timing of the intervention in the context of the collaboration.  Examples included decisions to intervene early in the task (i.e., while collaborators are still in ``exploratory mode'') and shortly after a search request (i.e., while collaborators' attention is focused on the system).  Finally, some factors relate to the status of the collaborator's ongoing conversation.  Examples included decisions to intervene when collaborators did not seem to be moving forward with the task (e.g., were silent or engaged in conversation unrelated to the task).

In Section~\ref{sec:discussion}, we discuss how these findings have design implications for systems that can take both dialog- and task-level initiative to support collaborative search.

\section{Related Work}
This research builds upon four areas of prior work: (1) collaborative search, (2) conversational search, (3) agent-supported collaborations, and (4) reference services.

\textbf{Collaborative Search:} Much prior work in collaborative search has developed standalone systems to support multiple searchers working together on an information-seeking task~\cite{Capra2012,Golovchinsky2009,Morris2006teamsearch,Morris2007,Morris2013,Paul2009,Putra2018,Shah2013,Yue2012,shah2010coagmento}. Such standalone systems typically include a search interface as well as additional tools and visualizations for collaborators to communicate, share information, and gain awareness of each other's activities.  Studies have found that standalone systems can help collaborators delegate tasks~\cite{Paul2009}, review each other's work~\cite{Htun2017,Capra2012}, avoid duplicating effort~\cite{Morris2007,Putra2018,Shah2013,Capra2012}, track their progress~\cite{Shah2013}, and improve the group's search performance~\cite{Capra2012}.

While standalone systems have been found to provide benefits, they have not been widely adopted by the general public~\cite{Morris2013,Hearst2014}. Morris~\cite{Morris2013} conducted a survey in which half the participants reported doing collaborative searches about once a week.  However, all participants reported using \emph{non-integrated} tools such as web search engines and communication tools such as email and instant messaging. Familiarity was the main motivator.  That is, participants preferred using tools that they already use as part of their daily routines.  Consequently, Morris proposed that future research should focus on embedding search technology into communication platforms that people already use to collaborate~\cite{Morris2013}.  As described next, a few studies have answered this call.

Hecht et al.~\cite{Hecht2012} introduced the SearchBuddies system, which was developed to automatically embed search results in response to questions posted on social media.  Study participants responded positively to the system \emph{except} when the embedded search results were relevant but also obvious. Closely related to our work, prior studies have also investigated \emph{searchbots} that can support collaborative searchers coordinating over an instant messaging platform (i.e., Slack)~\cite{avula2018searchbots,Avula2019}. \citet{Avula2019} presented a study in which pairs of participants could either: (1) only search individually, (2) only search directly from the Slack channel, or (3) both.  Participants reported greater awareness of their partner's activities when they could search directly from Slack (i.e., conditions 2 and 3).  However, participants also reported being distracted by their partner's interactions with the searchbot at times when they worked independently on different aspects of the task. \citet{avula2018searchbots} investigated searchbots that can monitor a conversation and \emph{proactively} intervene when they detect an information need.  The study compared two searchbot designs.  In one design, the searchbot elicited information before providing search results.  In another design, the searchbot automatically ``inferred'' the information need directly from the conversation.  Results found few differences between both designs.  Regardless of the design, participants perceived proactive interventions to be disruptive when they happened too soon (before participants understood the task requirements) and while participants were engaged in independent activities (i.e., ``divide-and-conquer'').

\textbf{Conversational Search:} The goal of conversational search is to develop systems that searchers can interact with using dialog. Radlinski and Craswell~\cite{Radlinski2017} enumerated five capabilities necessary for search systems to be considered ``conversational''. One of these is the ability for the system to engage in \emph{mixed-initiative} interaction.  That is, to be conversational, the system should be capable of taking control of the conversation to fulfill different objectives. Discourse analysis research has argued that task-oriented dialogs involve two levels of initiative: dialog- and task-level initiative~\cite{chu1997tracking}. Both levels of initiative place a discourse obligation on the other agent(s). However, they differ based on the agent's \emph{motivation} for taking initiative. Agents take \emph{dialog-level} initiative to establish mutual belief between agents. For example, taking dialog-level initiative might involve asking a clarification question in response to a request or recommendation.  On the other hand, agents take \emph{task-level} initiative when they wish to influence the goals of the other agents.  For example, taking task-level initiative might involve proposing a new approach to the task.

Developing a conversational search system begs the question: What should the system be capable of doing? Some research has aimed at defining the ``action space'' of a conversational search system. \citet{Vakulenko2019} argued that systems should be able to elicit information in response to a request and gather feedback about viable options. Such actions are examples of a system taking dialog-level initiative (i.e., establishing mutual belief). \citet{Azzopardi2018} proposed that systems should also be able to suggest options that may be relevant but have not been directly considered or requested. Such actions are examples of a system taking task-level initiative (i.e., making suggestions to influence the searcher's goals). 

Similar to our work, studies have used a Wizard of Oz methodology to better understand information-seeking conversations between \emph{pairs of humans}~\cite{vtyurina2017exploring,trippas2017people,budzianowski2018multiwoz,aliannejadi2019asking}. In these studies, one participant plays the role of the searcher and another participant (with access to a search engine) plays the role of the conversational search system. Vakulenko et al.~\cite{vakulenko2021large} analyzed different dialog datasets gathered using this methodology. Additionally, the authors included a
new dataset of virtual reference interviews between real-world library patrons and librarians.  Results found that reference librarians take high levels of initiative---more so than participants playing the role of the conversational search system in other datasets.

From the system side, prior work has primarily focused on conversational search systems that can take dialog-level initiative by asking clarification questions in response to a request.  Research has focused on predicting \emph{when} to ask for clarification~\cite{Arguello2017,zhang2020evaluating,christakopoulou2016towards} and \emph{which} clarification question(s) to ask in a specific context~\cite{aliannejadi2019asking,hashemi2020guided,rao2018learning,rao2019answer,sun2018conversational,zamani2020analyzing,zamani2020generating}.

\textbf{Agent-Supported Collaborations:} Embedding conversational agents in a collaboration has also been explored in areas such as intelligent tutoring~\cite{kumar2011conversational} and organizational behavior~\cite{larson2020leading}. Specifically, research on intelligent tutoring provides important insights about collaborations that are important to our research. For instance, to design the types of support that students need during collaborations, \citet{bales1950interaction} identified two key processes that happen during collaborations: instrumental (task-related) vs. expressive (socio-emotional). The former relates to activities that are primarily focused on the task's objective, and the later relates to activities primarily focused on the group's well-being. An imbalance of attention to one type of process tends to degrade the group's functionality. Studies have devised strategies around these processes that tutors can adopt to be more effective. For example, \citet{kumar2011conversational} suggest that in a group setting involving students and a tutor, tutors are generally pushed to the periphery and struggle to make the group consider their suggestions. To improve the involvement of tutors (or intelligent agents who may play the role of a tutor), research has explored conversational strategies that incorporate socio-emotional processes~\cite{ai2010exploring,kumar2010socially}. Some of these strategies include designing social prompts wherein an agent can check-in regarding the students' progress, provide encouragement when group members are inactive, or by actively listening to their discussions and encouraging their ideas. These studies found that students benefit from such conversational strategies---they help release tension in the group and help students reach consensus during discussions. While collaborative search is not identical to collaborative learning, our results also found that the Wizard was attentive to the social and emotional needs of collaborators in our study.

\textbf{Reference Services:} In our study, we decided to recruit reference librarians to play the role of the searchbot.  Reference librarians are trained to perform reference interviews with library patrons.  A reference interview is a mixed-initiative conversation in which the reference librarian helps the library patron resolve an information need~\cite{mabry2004reference}.  Reference librarians are trained to follow principles such as: (1) being forthcoming about knowledge gaps, (2) treating patrons as equal partners, and (3) treating each patron as a unique individual~\cite{mabry2004reference}. \citet{brooks1983using} argued that reference interviews are a useful resource to learn about the types of actions that search systems should try to emulate when supporting an individual searcher.  Our research is motivated by this idea but focuses on \emph{multiple} searchers working together.

\section{Methods}\label{sec:methods}

To investigate our two research questions, we conducted a Wizard of Oz lab study with 27 participant pairs (11 male, 33 female, and 10 did not specify).  All participants were undergraduate students at our university.  We wanted each pair of participants to already know each other (i.e., be friends or acquaintances).  Therefore, participants were enrolled in pairs.  We recruited three reference librarians (all female) to play the role of the ``Wizard''.  The study was approved by our university's Institutional Review Board (or Independent Ethics Committee). 

During the study, participants completed three collaborative search tasks.  The study used a within-subjects design and each pair of participants was exposed to three searchbot conditions (Section~\ref{sec:conditions}).  All three tasks were \emph{decision-making} tasks (Section~\ref{sec:tasks}).  During each task, participants had to consider different alternatives and make a joint selection.  Participants were seated in different rooms and had to communicate with each other (and the searchbot) using the Slack messaging platform.  To gather information, participants had to issue \emph{search requests} to the searchbot directly from the Slack channel. The Wizard, playing the role of the searchbot, sat in a third room.  The Wizard had access to the participant's Slack channel and could monitor the conversation.  Additionally, the Wizard used a custom-build web application to: (1) search the web, (2) send individual search results directly to the Slack channel, and (3) send clarification questions and/or task-level suggestions depending on the searchbot condition.  Participants could open and examine search results independently using their own individual browsers.  However, they were instructed to not search independently (e.g., using Google).  The Wizard's search interface was developed using the Bing Web Search API.

\subsection{Study Protocol and Design}\label{sec:protocol}

\begin{figure*}[t]
	\centering
	\begin{subfigure}[t]{0.45\textwidth}
		\centering
		\includegraphics[width =\columnwidth]{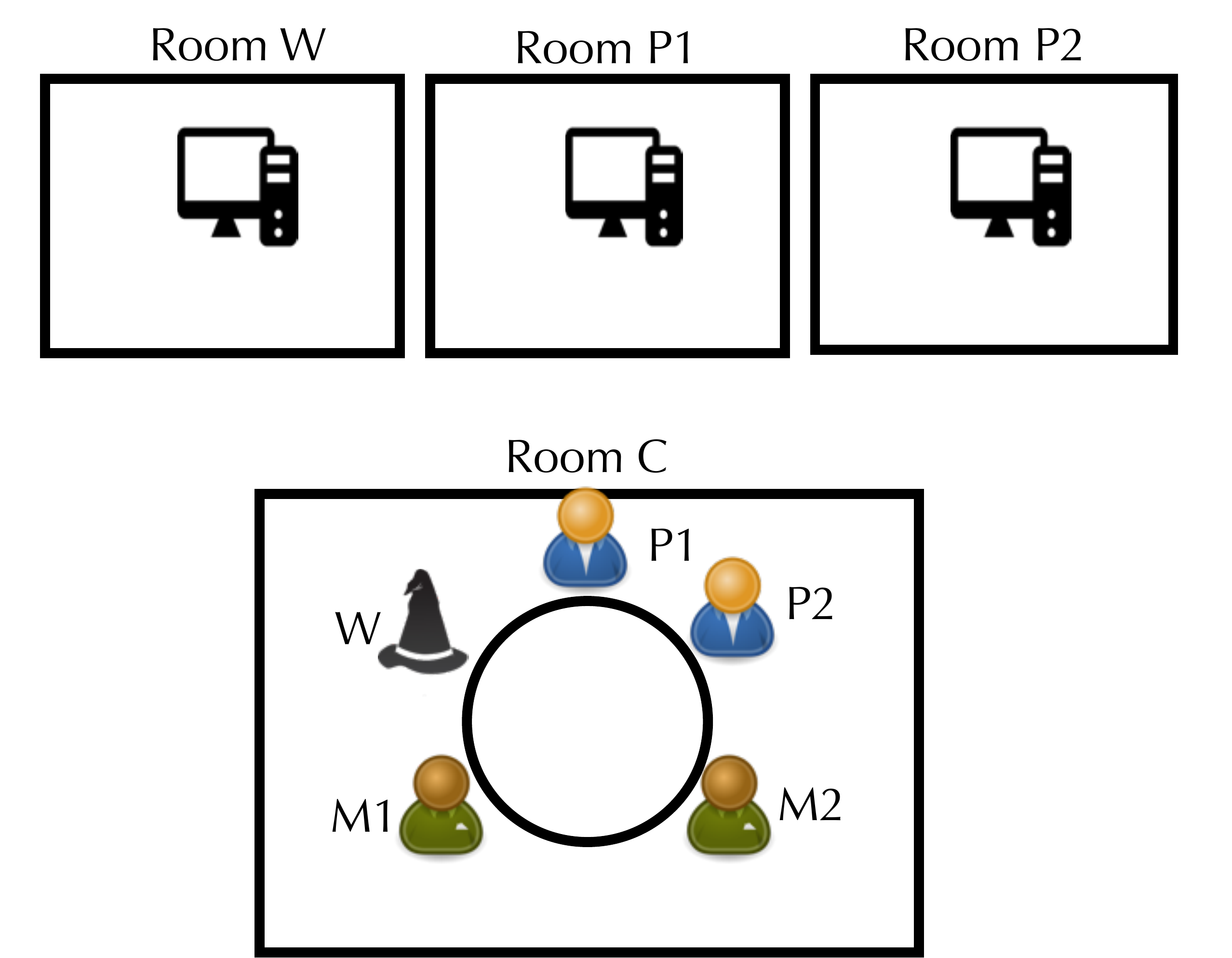}
		\caption{At the start of each study session, the lead moderator (M1) explained the purpose and protocol of the study to both participants (P1 and P2). Additionally, before each task, the lead moderator explained the next searchbot condition in the \emph{presence} of the Wizard (W). Participants could ask questions about the searchbot's capabilities in the next condition.}
		\label{subfig:intro}
	\end{subfigure}
	\hspace{1cm}
	\begin{subfigure}[t]{0.45\textwidth}
		\centering
		\includegraphics[width =\columnwidth]{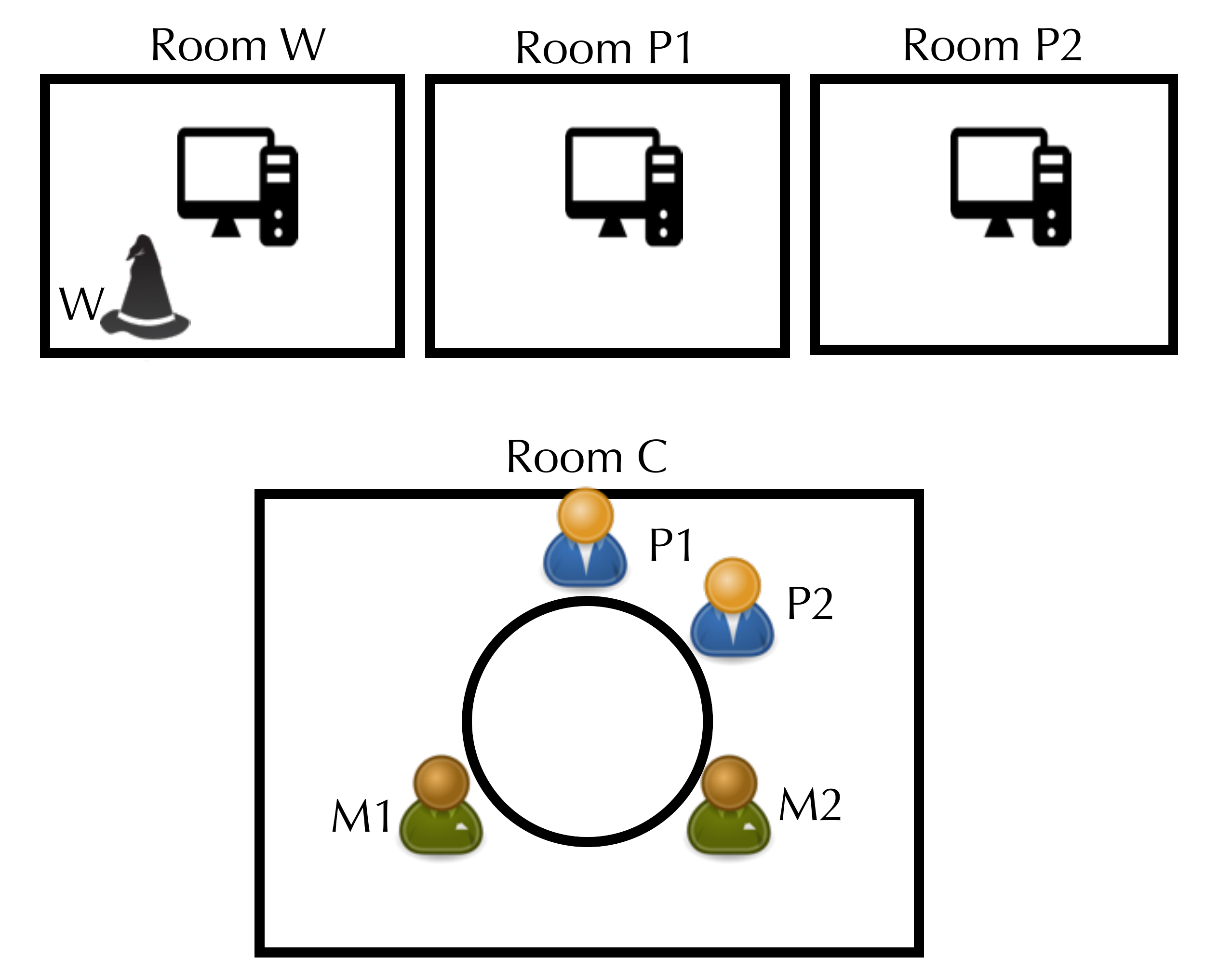}
		\caption{After explaining the next searchbot condition, the lead moderator read the next search task description aloud to the participants in the \emph{absence} of the Wizard. We did not want the Wizard to learn about specific criteria participants were asked to consider.}
		\label{subfig:task}
	\end{subfigure}	
	\begin{subfigure}[t]{0.45\textwidth}
		\centering
		\includegraphics[width =\columnwidth]{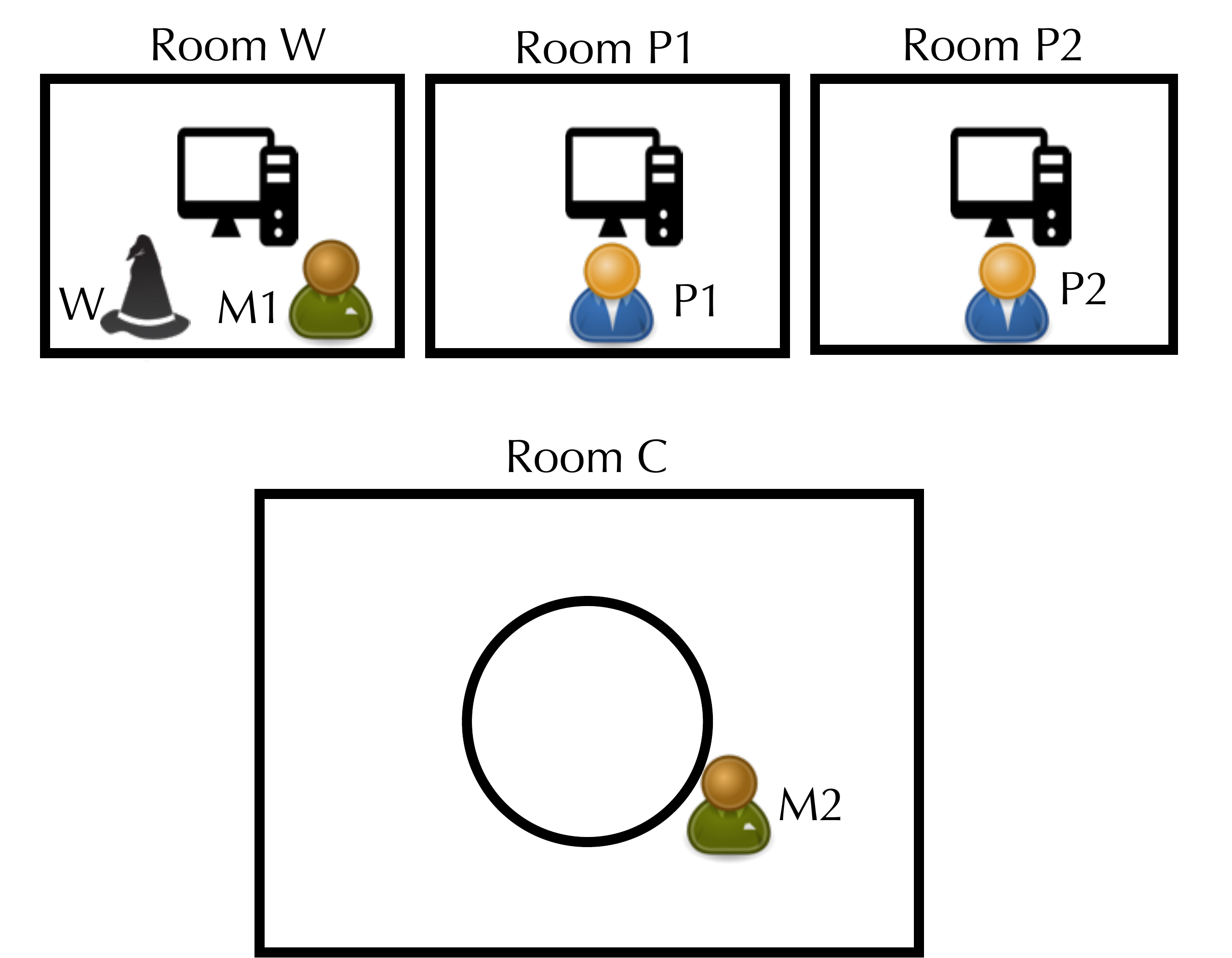}
		\caption{During each search task, both participants and the Wizard sat in separate rooms. The lead moderator sat behind the Wizard to assist with any technical issues.}
		\label{subfig:search}
	\end{subfigure}
	\hspace{1cm}
	\begin{subfigure}[t]{0.45\textwidth}
		\centering
		\includegraphics[width =\columnwidth]{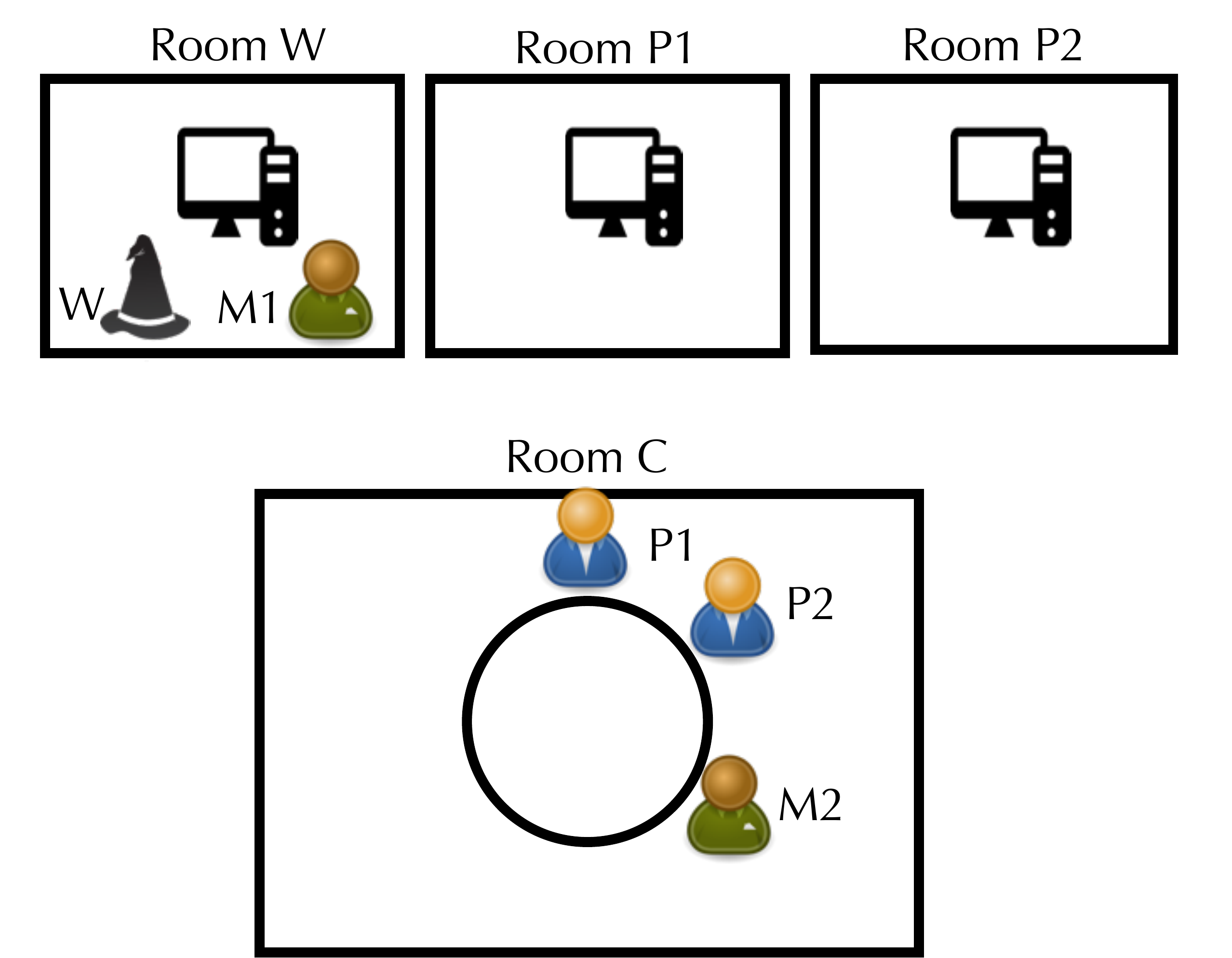}
		\caption{After each task, both participants were asked to verbally explain their solution to the task to the secondary moderator (M2). This exercise was done to discourage participants from satisficing.}
		\label{subfig:solution}
	\end{subfigure}
	\caption{Physical setup during different phases of the study session.}\label{fig:rooms}
\end{figure*}

\textbf{Study Protocol:} The study involved two moderators (M1 and M2) and took place in a physical space with four rooms: one common room (\textbf{Room C}), one room for the Wizard (\textbf{Room W}), and one room for each participant (\textbf{Room P1} and \textbf{Room P2}). Figure~\ref{fig:rooms} illustrates where members of the study were located during different phases of the study. 

The study had the following protocol.  At the start of the study session, both moderators welcomed the participants and the Wizard in \textbf{Room C} (Figure~\ref{subfig:intro}).  Here, the primary moderator (M1) described the purpose and protocol of the study.  Then, participants completed three tasks that followed the same sequence of steps.  First, the primary moderator explained the next searchbot condition.  As previously mentioned, our searchbot conditions varied based on the level of initiative the Wizard could take.  We explained each searchbot condition in the presence of the Wizard so that participants could ask any clarification questions directly to the Wizard.  Next, the Wizard left for \textbf{Room W} and the primary moderator read the next task description to the participants (Figure~\ref{subfig:task}).  Each participant was given a printed copy of the task description to keep during the task.  The Wizard was absent when participants were read the next task description. This was done to simulate a realistic scenario in which the ``system'' is unaware of precise details related to a searcher's objective.  After learning about the next task, participants went to their respective rooms (\textbf{Room P1} and \textbf{Room P2}).  Participants had 15 minutes to complete each task.  During the task, the primary moderator was present in \textbf{Room W} to help the Wizard with any technical issues (Figure~\ref{subfig:search}). After finishing the task, participants completed a post-task questionnaire.  After completing each task (i.e., search task + post-task questionnaire), participants were asked to explain their solution to the secondary moderator (M2) in \textbf{Room C} (Figure~\ref{subfig:solution}).  This was done to encourage participants to take the task seriously.  

Toward the end of the study, after all three tasks, the primary moderator conducted a stimulated-recall interview with the Wizard.  During this interview, the moderator and Wizard revisited \emph{every} instance in which the Wizard took initiative in the \textsc{BotDialog} and \textsc{BotTask} condition.  For each instance, the moderator asked the Wizard to comment on: (1) their motivation for taking initiative and (2) their rationale for deciding that the timing of the intervention was appropriate.  Each participant received US\$20 for participating in the study and the Wizard received US\$30 per study session.

\textbf{Study Design:} Our study involved three searchbot conditions and three search tasks.  Each participant pair was exposed to all three searchbot conditions and all three search tasks.  We wanted to balance the order in which participants were exposed to searchbot conditions and search tasks.  A Latin square with three treatments (i.e., A, B, C) involves three treatment orders (i.e., ABC, CAB, BCA), which ensures that each treatment appears exactly once in each position.  To accommodate our two factors (i.e., searchbot condition and search task), we used the cross-product of two Latin squares, which yielded 9 orderings (i.e., 3 searchbot condition orders $x$ 3 search task orders = 9 orderings).  Each of these 9 orderings was completed by exactly 3 participant pairs (i.e., hence 27 study sessions).  For example, using letters to denote searchbot conditions and numbers to denote search tasks, 3 participants were exposed to ordering A1, B2, C3; 3 participants were exposed to ordering B3, C1, A2; and so on.  Finally, as previously mentioned, our study involved 3 reference librarians playing the role of Wizard.  Each of the 9 orderings was completed by all 3 reference librarians (i.e., 9 study sessions per librarian).


\subsection{Searchbot Conditions}\label{sec:conditions}

Each pair of participants experienced three searchbot conditions (i.e., a within-subjects design). In this paper, we report on the Wizard's decisions to take initiative in the \textsc{BotDialog} and \textsc{BotTask} conditions.  Therefore, in this section, we only describe these two conditions. In the other condition (called \textbf{\textsc{BotInfo})}, the Wizard could only respond with a search result in response to each search request.

\textbf{\textsc{BotDialog}:} In this condition, the searchbot could take dialog-level initiative but not task-level initiative.  Agents take dialog-level initiative to establish mutual belief between agents~\cite{chu1997tracking}. Specifically, in this condition, the Wizard could take dialog-level initiative by asking one or more clarification questions in response to a search request. As in all searchbot conditions, participants issued search requests using the `@Wizard' preamble (e.g., ``@Wizard how safe is Karongwe''). In response to a request, the Wizard could ask one or more clarification questions in order to better understand the participants' needs. After asking one or more clarification questions, the Wizard sent a single search result to the participants' Slack channel. After sending a search result, the Wizard could not ask any more clarification questions about the request. The Wizards were instructed to ask clarification questions when additional information might help them find a better search result to satisfy the participants' needs. Many of these clarification questions were asked when participants issued search requests that were too ambiguous or broad. As previously mentioned, recent research has mostly focused on developing conversational search systems that can ask clarification questions in response to a request~\cite{aliannejadi2019asking,hashemi2020guided,rao2018learning,rao2019answer,sun2018conversational,zamani2020analyzing,zamani2020generating}.  Hence, the \textsc{BotDialog} condition was meant to represent conversational search systems we may see in the near future.

\textbf{\textsc{BotTask}:} In this condition, the Wizard could take both dialog- and task-level initiative. Agents take task-level initiative when they explicitly intend to influence the other agents' goals~\cite{chu1997tracking}.  In this condition, as in the \textsc{BotDialog} condition, the Wizard could ask one or more clarification questions in response to a search request. Additionally, the Wizard could also provide task-level suggestions. In other words, the Wizard could lead the conversation by providing suggestions with the intent to influence the participants' goals or approach to the task~\cite{chu1997tracking}. The Wizard could provide task-level suggestions either in response to a search request or by \emph{proactively} intervening in the conversation. The Wizard was instructed to monitor the conversation over Slack and intervene if they thought they could provide valuable task-level advice. The \textsc{BotTask} condition was meant to represent conversational search systems we may see farther into the future, which may be able to proactively influence the goals and strategies of users during a search task.

\subsection{Tasks}\label{sec:tasks}

During the study, participants completed three collaborative search tasks.  All three tasks were \emph{decision-making} tasks, which involve comparing alternatives along different dimensions or criteria and make a joint selection based on the information gathered.  Our tasks left the alternatives open-ended but specified the criteria that participants should consider.  Each task involved three specific criteria.  To encourage participants to take the task seriously, participants were asked to justify their final selection to the secondary moderator after finishing the task.  Our three tasks involved the following themes: (1) volunteer planning, (2) vacation planning, and (3) dinner planning.  Each task was contextualized using a background scenario.  Participants were assigned gender-neutral names (i.e., Jamie and Alex) during the study.  As an example, the volunteer planning task had the following background scenario and objective statement:

\textbf{Background:} Jamie and Alex are rising seniors. They have decided that before they finish school, they would like to spend a summer volunteering (2-3 months). Jamie heard from a friend that volunteering internationally is an option they should consider. They have decided to explore different volunteering programs in Africa.

\textbf{Objective:} Jamie mentioned to Alex that their parents would only agree to a volunteering program in Africa if they are confident about their safety, the affordability of the plan, and if the project is exciting. In this task, with the help of the searchbot, work together to find a project that suits both of you.

Our tasks specified the following three criteria for participants to consider: (1) volunteer planning---(a) safety, (b) affordability, and (c) interestingness; (2) vacation planning---(a) things to do/see, (b) ease of obtaining a visa, and (c) safety; and (3) dinner planning---(a) type of cuisine, (b) difficulty level, and (c) possible allergens to avoid.

\subsection{Wizard Training Session}\label{sec:wizard_training}

While pilot testing the study protocol, we realized that the three reference librarians might benefit from a training session.  During the training session, we explained the purpose of the study and the three searchbot conditions.  We also introduced the three search tasks without disclosing exact details about the participants' objectives.  For each task, the reference librarians were asked to brainstorm ways in which they might assist participants by asking clarification questions and/or making task-level suggestions.  During the study, the reference librarians were given copies of these notes to help them remember ways to help the participants.

\subsection{Stimulated Recall Interview}\label{subsec:stimulated_recall}

Toward the end of each study session, the lead moderator conducted a stimulated recall interview with the Wizard. During this interview, the moderator used the chatlog captured over the Slack channel. To gain insights about the Wizard's actions, the moderator revisited every instance in the \textsc{BotDialog} and \textsc{BotTask} condition in which the Wizard asked a clarification question or made a task-level suggestion. For each intervention, the moderator asked two questions: (RQ1) What were you hoping to achieve and why? and (RQ2) Why did you decide this would be an appropriate time to intervene? The Wizard's responses were audio recorded and analyzed using qualitative techniques.

\begin{figure*}[t]
	\centering
	\begin{subfigure}[t]{0.48\textwidth}
		\centering
        \includegraphics[width =\columnwidth]{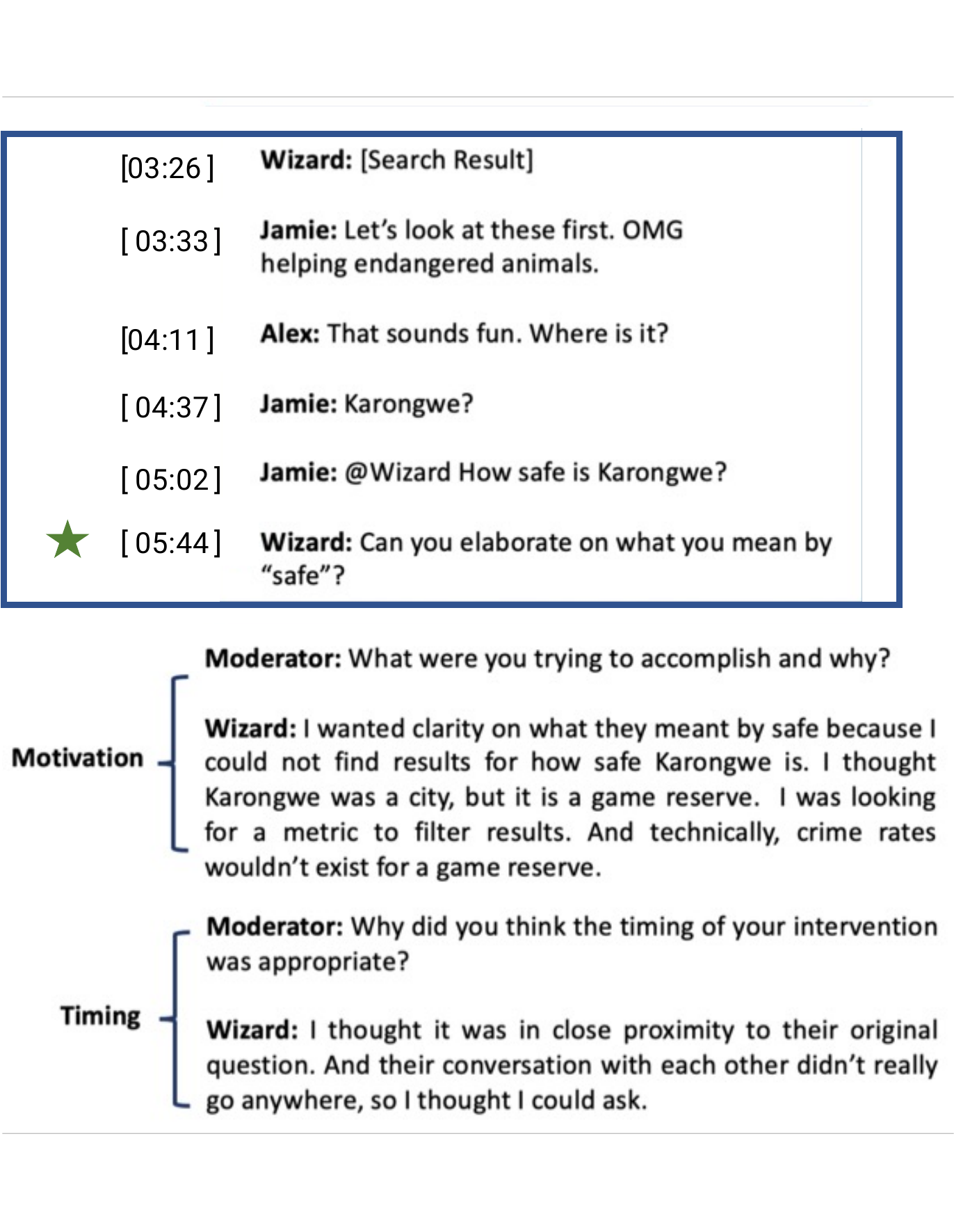}
		\caption{\textsc{BotDialog}: In this example, the Wizard is asked about their decision to intervene in order to ask a clarification question in response to a search request (i.e., the meaning of ``safe'').}
		\label{subfig:botdialog_sr}
	\end{subfigure}
	\hspace{0.1cm}
	\begin{subfigure}[t]{0.48\textwidth}
		\centering
        \includegraphics[width =\columnwidth]{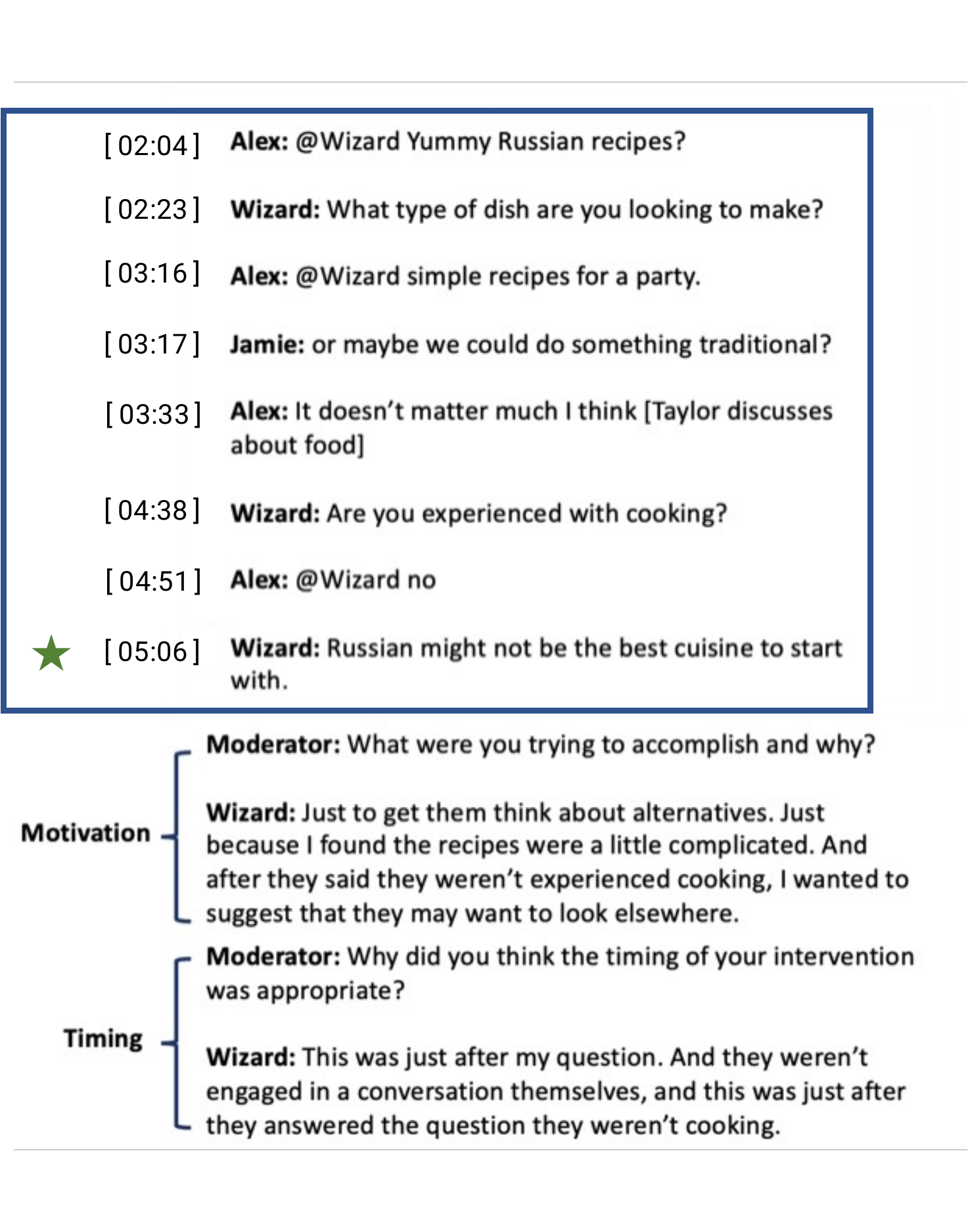}
		\caption{\textsc{BotTask}: In this example, the Wizard is asked about their decision to proactively intervene the with a task-level suggestion (i.e., to suggest that an option being considered might not be appropriate).}
		\label{subfig:bottask_sr}
	\end{subfigure}
	\hspace{1cm}
	\caption{Examples of the \textsc{BotDialog} and \textsc{BotTask} conditions showing how the Wizards were asked to reflect on their actions during the stimulated recall interview. The timestamps correspond to the minutes and seconds [MM:SS] from the start of the search task.}\label{fig:sr}
\end{figure*}

To further elaborate on the interview process, consider the examples in Figure~\ref{subfig:botdialog_sr} (\textsc{BotDialog}) and Figure~\ref{subfig:bottask_sr} (\textsc{BotTask}).  As illustrated in Figure~\ref{subfig:botdialog_sr}, the moderator asked about an instance (denoted by $\star$) in which the Wizard intervened to ask a follow-up question in response to a search request.  Here, in response to the question about their motivation, the Wizard stated that they needed clarification on what the participants meant by ``safe'' (i.e., crime rate vs.~dangerous wildlife).  In response to the question about the timing of the intervention, the Wizard stated that they intervened shortly after a search request and that the participants were not actively engaged in conversation (i.e., were idle).  As illustrated in Figure~\ref{subfig:bottask_sr}, the moderator asked about an instance in which the Wizard intervened to make a task-level suggestion.  In response to the question about their motivation, the Wizard stated that they wanted participants to know that an option being considered (i.e., Russian cuisine) might not be compatible with a previously stated preference (i.e., recipes for inexperienced cooks).  In response to the question about the timing of the intervention, the Wizard stated that the intervention was in close proximity to a previous intervention (i.e., the participants attention was probably focused on the Wizard).

\subsection{Data Analysis}
We conducted inductive qualitative coding \citep{levitt2018journal} to analyze the Wizard's responses to open-ended questions during the post-task interview. We took an iterative, consensus approach to coding \citep{levitt2018journal, hill1997guide}, meaning that we conducted multiple iterations of coding to develop and refine the themes, and resolved differences from individual coding until the full agreement was achieved. We identified themes (and subthemes) that characterize: (\textbf{RQ1}) the Wizard's motivations for intervening and (\textbf{RQ2}) the Wizard's rationales for the timing of each intervention. The Wizard's responses during the stimulated recall interview were prepared and analyzed as follows. In terms of data preparation, we first transcribed the audio recordings of all interviews. Second, we aligned the Wizard's comments with the corresponding intervention in the chatlog captured over the Slack channel. This was done to provide necessary context of the intervention during the data analysis process. In terms of data analysis, two of the authors coded the data in an iterative fashion. During each iteration, both authors independently coded the same subset of data. The authors independently developed codes (i.e., subthemes) associated with the Wizard's motivations for intervening and rationales for the timing of an intervention. After this, both authors met to merge their codes, clarify code definitions, and resolve any disagreements. Then, the authors moved on to the next iteration (i.e., another subset of the data). When a new code was added, both authors re-visited all previously coded data to apply the new code. Ultimately, both authors coded 100\% of the data and reached full consensus regarding the application of all codes. Finally, both authors organized the codes/subthemes into higher-level themes.

\subsection{Decisions and Rationale}\label{subsec:rationales}
In designing our study, we made a few decisions that might benefit from justification.

\textbf{Participants Knew the Searchbot was Human:} Typically, in a ``Wizard of Oz'' study, participants interact with a ``computer system’’ that is actually controlled by a human~\cite{Martin2012}. In our study, participants were fully aware that the searchbot was a human reference librarian. We made this decision for two reasons.  First, we wanted participants to have high and consistent expectations about the searchbot’s abilities across conditions and study sessions.  Research has found that expectations strongly influence engagement with an AI system~\cite{Khadpe2020,Luger2016}. Second, providing task-level advice (and doing so proactively) goes far beyond the current capabilities of commercial systems such as Siri and Alexa.  Thus, we found it unlikely for participants to believe they were interacting with a fully automated system in the \textsc{BotTask} condition.

\textbf{Participants Knew About the Searchbot's Capabilities Before Each Task:} As previously mentioned, before each task, participants were explained the capabilities of the searchbot in the next condition.  This was done to set expectations from the outset of the task. Participants had 15 minutes to complete each task.  Therefore, we did not want them to spend part of this time figuring out the capabilities of the searchbot on their own.

\textbf{The Searchbot Returned a Single Web Result Per Request:} In all conditions, the searchbot processed search requests by returning a \emph{single} search result versus a ranked list.  This decision was made to encourage participants to engage with the searchbot.  We believe that providing a ranked list would have significantly reduced participants’ engagement with the searchbot.

\textbf{Number of Reference Librarians:} Deciding on the number of references librarians was another crucial decision. On one hand, we did not want to choose a single reference librarian for the whole study.  We did not want our results to be heavily influenced by the behavior of a single individual.  On the other hand, recruiting and training 27 reference librarians was not feasible. Ultimately, we decided to recruit three reference librarians.  Each reference librarian played the role of Wizard during 9 study sessions. As previously noted, we used the cross product of two Latin squares to vary the order of the searchbot condition and search task, resulting in 9 orderings.  Each reference librarian experienced all 9 orderings, resulting in 27 study sessions.

\section{Results --- RQ1: Motivations}\label{sec:motivations_timing}
In this section, we present (sub)themes that characterize the Wizard's motivations for taking initiative. These (sub)themes were derived from the Wizard's responses to the question: ``During this intervention, what were you hoping to achieve and why?'' As previously mentioned, the Wizard could take different levels of initiative during the \textsc{BotDialog} and \textsc{BotTask} conditions. During the \textsc{BotDialog} condition, the Wizard could only ask clarification questions in response to a search request. During the \textsc{BotTask} condition, in addition to asking clarification questions, the Wizard could also provide task-level suggestions, either in response to a request or by proactively intervening. For each subtheme, we present the most illustrative example. We use \textbf{[D]} and \textbf{[T]} to denote the searchbot condition (dialog- and task-level initiative) where a subtheme occurred and use S01-S27 to denote the study session associated with each example. The timestamps correspond to the minutes and seconds [MM:SS] from the start of the search task.

\subsection{Theme: Common Ground}
Not surprisingly, the Wizard intervened when they wanted to better understand the participants' goals. We identified three different scenarios associated with this motivation: (1) when the Wizard wanted to learn about the participants' \emph{problem space} (e.g., task-related constraints); (2) when the participants expressed a constraint or personal preference using vague and subjective language; and (3) when the Wizard could not follow the participants' conversation to make accurate inferences about options under \emph{active} consideration.

\subsubsection{\textbf{Problem Information [D\&T]}}
During each task, participants compared different options and made a final selection based on their goals and constraints. However, participants did not always express those elements explicitly in their search requests. In such cases, the Wizard asked for additional information about those goals and constraints. Gathering this information helped the Wizard understand the context of the search and narrow the search space. Consider the following example from S01 (\textsc{BotDialog)}:

{\footnotesize 
\begin{itemize}[leftmargin=.05\linewidth,rightmargin=.05\linewidth]
\itemsep 0em 
\item[] \texttt{[01:41] Jamie: @Wizard Trips to Rio in December}	 
\item[] \texttt{[01:58] Wizard: What is this trip for?}	 
\end{itemize}
}

\emph{Wizard's comments:} \emph{``...the purpose of what they're looking for. They didn't seem to know what they were looking for so I thought looking at the purpose might help direct them a little bit.''} 

\subsubsection{\textbf{Metrics for Evaluation [D\&T]}}
During each task, participants compared options along different criteria. For example, when comparing travel destinations, participants considered dimensions such as safety, affordability, and visa requirements. In some cases, participants communicated their preferences along these dimensions using subjective terms (e.g., `cheap', `easy'). Such qualifiers are problematic because they mean different things to different people. In such cases, the Wizard followed-up either by: (1) asking participants to express their preferences using a standard metric system or (2) asking participants to elaborate using more concrete language. To illustrate, consider the following example from S16 (\textsc{BotDialog}):

{\footnotesize 
\begin{itemize}[leftmargin=.05\linewidth,rightmargin=.05\linewidth]
\itemsep 0em 
\item[] \texttt{[02:18] Jamie: @Wizard what countries in Africa experience moderate climates in the summer?}	 
\item[] \texttt{[06:50] Wizard: Do you want temperatures that are in a certain range? How do you want to measure climate?} 
\end{itemize}
}

\emph{Wizard's comments:} \emph{``They wanted me to search somehow by temperature. Words like moderate or best climate are ambiguous.''}

\subsubsection{\textbf{Lost Track of Options/Criteria [D\&T]}}
During each task, the Wizard could monitor the conversation to keep track of the different options and evaluation criteria being considered. In some cases, however, the Wizard lost track of the conversation and decided to take initiative to verify options and criteria under \emph{active} consideration. In the \textsc{BotDialog} condition, the Wizard elicited this information by asking a clarification question. In the \textsc{BotTask} condition, the Wizard proactively intervened to elicit this information or suggest a possible direction for participants to pursue. Consider the following example from S20 (\textsc{BotTask}). Here, the participants were considering Peru as a possible travel destination and quickly \emph{switched} to Argentina. The Wizard was uncertain if they wanted to actively pursue Argentina as a new option, so they intervened:

{\footnotesize 
\begin{itemize}[leftmargin=.05\linewidth,rightmargin=.05\linewidth]
\itemsep 0em 
\item[] \texttt{[01:11] Alex: what are the travel advisories for peru?}
\item[] \texttt{[02:38] Wizard: [Search Result]}
\item[] \texttt{[03:23] Jamie: it says don't go to some parts of cusco bc of terrorism}
\item[] \texttt{[03:41] Alex: is that the machu pichu place. terrorists be ruining things for us smh}
\item[] \texttt{[03:55] Wizard: there are demonstrations that are happening around peru, so you may want to avoid it until further notice}
\item[] \texttt{[04:09] Jamie: Argentina and Chile look to be safe}
\item[] \texttt{[04:25] Alex: should we look into Argentina? we can see messi}
\item[] \texttt{[05:47] Wizard: want to know more about Argentina and its travel advisories?}
\end{itemize}
}

\emph{Wizard's comments:} \emph{``They did mention Argentina. And it seemed like Peru, up to this point, kind of felt like a dead end. And so they wanted to look into Argentina... so I figured I'd just ask.''}

\subsection{Theme: Poor Search Request}

Sometimes, the Wizard decided that a search request was \emph{poor} (e.g., too ambiguous or broad). In such cases, the Wizard elicited additional details to help them find information that would better satisfy the participants' needs. Interestingly, the Wizard also intervened when the search request included constraints that would be challenging to address.

\subsubsection{\textbf{Ambiguous or Underspecified Request [D\&T]}}

The Wizard asked follow-up questions when the participants issued a search request that was too ambiguous, vague, or underspecified. Interestingly, in some cases, the Wizard made this determination using \emph{pre-retrieval} evidence (i.e., their own intuition). In other cases, they used \emph{post-retrieval} evidence (i.e., no useful search results returned). For the case of using \emph{pre-retrieval} evidence, consider the following example from S19 (\textsc{BotDialog}). Here, the participants requested information about activities in the city of Medellin. The Wizard immediately recognized the request as being too broad. Thus, the Wizard followed-up to elicit information about the \emph{types} of activities participants were interested in:

{\footnotesize 
\begin{itemize}[leftmargin=.05\linewidth,rightmargin=.05\linewidth]
\itemsep 0em 
\item[] \texttt{[06:25] Jamie: @Wizard What are activities to do in Medellin?}
\item[] \texttt{[06:42] Wizard:	What kind of activities do you have in mind?}
\end{itemize}
}

\emph{Wizard's comments:} \emph{``There're many different kinds of activities. I wanted to narrow it down to something so that I can search better.''}

For the case of using \emph{post-retrieval} evidence, consider the following example from S24 (\textsc{BotDialog)}. Here, the Wizard learned about the ambiguity associated with the concept of `safety' (e.g., crime rate vs.~dangerous wildlife) after viewing the search results:

{\footnotesize 
\begin{itemize}[leftmargin=.05\linewidth,rightmargin=.05\linewidth]
\itemsep 0em 
\item[] \texttt{[05:02] Jamie: @Wizard How safe is Karongwe?}
\item[] \texttt{[05:44] Wizard: Can you elaborate on what you mean by 'safe'?}
\end{itemize}
}

\emph{Wizard's comments:} \emph{``I thought Karongwe was a city [but it] was the name of a Game Reserve and so technically crime rates wouldn't really [apply] in a Game Reserve.''}

\subsubsection{\textbf{Challenging Constraints [D\&T]}}

In some cases, it was difficult for the Wizard to address all the constraints specified in a search request. This led the Wizard to negotiate with the participants on the constraints that were important to them. To illustrate, consider the following example from S27 (\textsc{BotDialog}): 

{\footnotesize 
\begin{itemize}[leftmargin=.05\linewidth,rightmargin=.05\linewidth]
\itemsep 0em 
\item[] \texttt{[09:52] Jamie: @Wizard between Uruguay and Chile, which has better nightlife in December?}
\item[] \texttt{[10:35] Wizard: Does it need to be in December specifically or can it be year-round?}
\end{itemize}
}

\emph{Wizard's comments:} \emph{``I figured the month would probably be the biggest factor that was throwing off the search results. They provided that extra specification without necessarily realizing that it was going to affect the whole search.''}

\subsection{Theme: Strategic Support}

The Wizard offered different types of strategic support by asking clarification questions and providing suggestions.

\subsubsection{\textbf{Filtering [D\&T]}} 

In many cases, the Wizard requested specific information to filter the search results. This process was a natural strategy, as each task had constraints that participants had to accommodate in their final selection. Thus, asking questions about specific task constraints helped the Wizard narrow the search space. To illustrate, consider the following example from session S13 (\textsc{BotDialog}):

{\footnotesize 
\begin{itemize}[leftmargin=.05\linewidth,rightmargin=.05\linewidth]
\itemsep 0em 
\item[] \texttt{[04:09] Jamie: @Wizard What are some easy middle eastern dishes to make?}	 
\item[] \texttt{[04:45] Wizard: Are there ingredients you want to use or to avoid?}	 
\end{itemize}
}

\textbf{Wizard's comments:} \emph{``I was hoping to gain just like any information that could help me narrow down [...] just like the search was pretty broad. And I didn't have a lot of other information to go on.''}

\subsubsection{\textbf{Make the Task More Manageable [D\&T]}} 

During each task, participants had to build consensus over multiple options and criteria. Sometimes, the Wizard observed participants not having an action plan on how to do this. In such cases, the Wizard intervened to help decompose the task into smaller subtasks to make the task more manageable. In the \textsc{BotDialog} condition, they did this by asking a series of follow-up questions, which \emph{indirectly} suggested ways to decompose the task. In the \textsc{BotTask} condition, they proactively interjected to give advice on how to decompose the task. To illustrate, consider the following example from S11 (\textsc{BotTask)}: 

{\footnotesize 
\begin{itemize}[leftmargin=.05\linewidth,rightmargin=.05\linewidth]
\itemsep 0em 
\item[] \texttt{[00:58] Jamie: @Wizard tourism destinations in brazil}
\item[] \texttt{[01:15] Wizard: What kind of activities are you interested in?}	 
\item[] \texttt{[01:32] Jamie: beaches, hiking}
\item[] \texttt{[01:54] Alex: @Wizard outdoor adventure activities}
\item[] \texttt{[03:03] Jamie: @Wizard temperature in Rio de Janeiro in december}
\item[] \texttt{[03:04] Wizard: [Search Result] you might also want to think about whether you just want to go to one city or to a few different places in the country}
\item[] \texttt{[04:02] Jamie: lets go like all around brazil not just in Rio}	 
\item[] \texttt{[04:29] Alex: the weather is 85 so pretty warm}	
\item[] \texttt{[04:35] Wizard: It'd help to plan to find a few locations and look up travel options between them.}
\item[] \texttt{[04:48] Jamie: let's say 3 locations? @Wizard top 3 places to visit in Brazil}	
\end{itemize}
}

\emph{Wizard's comments:} \emph{``...hoping to provide guidance about how they could approach making choices or decide what to search for next.''}

\subsubsection{\textbf{Converging Decisions [D\&T]}} 

Sometimes the Wizard noticed that the participants issued a search request that was \emph{inconsistent} with a previously made decision that could impact the search. In such situations, the Wizard intervened to verify whether they should consider the previously made decision as a global parameter and incorporate it into the current search. For example, in S11 (\textsc{BotDialog)}, the participants had decided to make pho just before issuing a search request about a new dimension of the task (i.e., allergen). In response to this request, the Wizard asked whether they should focus the search on allergens found in pho: 

{\footnotesize 
\begin{itemize}[leftmargin=.05\linewidth,rightmargin=.05\linewidth]
\itemsep 0em 
\item[] \texttt{[00:54] Jamie: Bro Vietnam is where it is at. Let's make pho.}
\item[] \texttt{[01:23] Jamie: @Wizard what food are people most allergic to?}
\item[] \texttt{[02:13] Wizard: Food people are allergic to that would be in Pho?}
\end{itemize}
}

\emph{Wizard's comments:} \emph{``They hadn't had any conversation specifically about Pho before, then they asked a more generic question. So I wanted to see if they wanted to combine those strains. It seemed like they're interested in combining those streams of questioning.''}

\subsubsection{\textbf{Provide Alternatives [T]}} 

In the \textsc{BotTask} condition, the Wizard sometimes proactively suggested alternatives that the participants had not discussed but should consider based on criteria mentioned in the conversation. This happened when the Wizard thought participants needed to re-evaluate their search strategy or needed more ideas. To illustrate, consider the example from S01 (\textsc{BotTask}). Here, the Wizard suggested a cuisine that participants had not considered but was consistent with specific preferences and constraints mentioned in the conversation (i.e., international and appropriate for an inexperienced cook): 

{\footnotesize 
\begin{itemize}[leftmargin=.05\linewidth,rightmargin=.05\linewidth]
\itemsep 0em 
\item[] \texttt{[01:19] Alex: russian?? like everyone knows russia but no one knows the food options}
\item[] \texttt{[02:04] Jamie: @Wizard yummy russian recipes}
\item[] \texttt{[02:23] Wizard: what type of dish are you looking to make?}
\item[] \texttt{[03:00] Alex: idk what kind of food russia has haha soup?}
\item[] \texttt{[03:16] Jamie: @Wizard simple recipes for a party}
\item[] \texttt{[04:03] Wizard: [Search Result]}
\item[] \texttt{[04:38] Wizard: Are you experienced in cooking?}
\item[] \texttt{[04:51] Jamie: @Wizard no}
\item[] \texttt{[05:06] Wizard: Russian might not be the best cuisine to start with.}
\item[] \texttt{[06:06] Jamie: @Wizard what are some international dishes that are easy to make for groups?}
\item[] \texttt{[06:46] Wizard: do you have have preferences about cuisines you DON'T want to look at?}
\item[] \texttt{[07:16] Jamie: @Wizard italian, chinese, mexican, american and japanese}
\item[] \texttt{[08:25] Wizard: How about mediterranean?} 
\end{itemize}
}

\emph{Wizard's comments:} \emph{``I asked what they did not want and they gave a list of really popular cuisines... also, going back, they said they aren't experienced with cooking. So I was just thinking from my own experience what kind of cuisines are fairly easy to cook.''}

\subsubsection{\textbf{Lacking Focus [D\&T]}} 

In some cases, the participants struggled to narrow their options down to a manageable set. In the \textsc{BotDialog} condition, the Wizard asked clarification questions to explicitly help the participants reduce the search space based on specific dimensions (e.g., ``Are there ingredients you want to use or avoid?''). Importantly, based on the Wizard's comments, these interventions served two goals: (1) to narrow the current search request and (2) to provide the participants with a specific direction on how to approach the task in a more focused manner. This ultimately helped the participants overcome indecision and move forward with the task. In the \textsc{BotTask} condition, the Wizard took a more proactive approach. For example, in S16 (\textsc{BotTask}), the Wizard noticed that the participants could not focus on a single activity (being indecisive), and thus decided to intervene:

{\footnotesize 
\begin{itemize}[leftmargin=.05\linewidth,rightmargin=.05\linewidth]
\itemsep 0em 
\item[] \texttt{[11:12] Jamie: did it tell you the price of the kayaking}
\item[] \texttt{[11:17] Alex: camping kayaking and like hiking}
\item[] \texttt{[11:22 - 11:42] [Jamie and Alex keep discussing different outdoor activities without settling]}
\item[] \texttt{[12:07] Wizard: We could look more into options for a specific activity, such as hiking.}
\end{itemize}
}

\emph{Wizard's comments}: \emph{``The conversation hadn't settled on something they definitely wanted to do... it seemed like focusing on one particular thing might be a good strategy for them at that point.''}

\subsubsection{\textbf{Lacking Direction [D\&T]}}

In contrast to the previous subtheme, in some cases, participants seemed to lack important knowledge needed to even understand the ``option space''. Similar to the previous subtheme, in the \textsc{BotDialog} condition, the Wizard asked clarification questions to \emph{indirectly} provide this necessary knowledge. In the \textsc{BotTask} condition, the Wizard took a more proactive approach. For example, in S25 (\textsc{BotTask}), the participants asked about ``visa requirements for South America''. Based on this request (and the ongoing conversation), the Wizard determined that participants might not know that visa requirements vary by country of destination and the traveler's country of citizenship: 

{\footnotesize 
\begin{itemize}[leftmargin=.05\linewidth,rightmargin=.05\linewidth]
\itemsep 0em 
\item[] \texttt{[00:37] Jamie: @Wizard what's the process for getting a visa for south america? I know nothing about the process} 
\item[] \texttt{[00:51] Alex: honestly me neither}
\item[] \texttt{[01:24] Wizard: It varies depending on what country you're going to and what country you have citizenship from.}
\end{itemize}
}

\emph{Wizard's comments:} \emph{``There were so many variables in it [the question]... the question was so general they didn't seem to know where they're going... I was like, this is the best way to respond.''}

\subsection{Theme: Auxiliary Support} 

In the \textsc{BotTask} condition, the Wizard engaged with participants in different ways beyond asking clarification questions. Some of these interventions were motivated by the Wizard wanting to: (1) provide participants with additional information about a shared search result; (2) inform participants about challenges encountered during a search; or (3) check-in with participants and ask whether they needed further assistance with the task. 

\subsubsection{\textbf{Justifying Result [T]}}

In the \textsc{BotTask} condition, the Wizard sometimes accompanied the shared result with a justification when they thought the additional information would help participants understand the relevance of the result. To illustrate, consider the following example from S20 (\textsc{BotTask}): 

{\footnotesize 
\begin{itemize}[leftmargin=.05\linewidth,rightmargin=.05\linewidth]
\itemsep 0em 
\item[] \texttt{[Jamie and Alex talk about options for outdoor recreation for a while.]}
\item[] \texttt{[10:52] Jamie: there are nature tours too.}
\item[] \texttt{[12:09] Alex: i think they have a famous opera house or something? or maybe it was a theatre}
\item[] \texttt{[12:30] Jamie: YUH they do. it's a rly cool place}
\item[] \texttt{[12:55] Wizard: Since you were interested in some outdoor activities, I have sent a webpage with a list for outdoor activities.}
\end{itemize}
}

\emph{Wizard's comments:} \emph{``I wanted to provide them some extra information based off what they were initially talking about. I wasn't sure if they were still with the outdoor activities so I added [the reasoning].''}

\subsubsection{\textbf{Inform Challenges [T]}}

In the \textsc{BotTask} condition, the Wizard sometimes provided task-level suggestions that were motivated by challenges they faced while searching on behalf of the participants. Consider the following example from S17 (\textsc{BotTask}): 

{\footnotesize 
\begin{itemize}[leftmargin=.05\linewidth,rightmargin=.05\linewidth]
\itemsep 0em 
\item[] \texttt{[09:32] Jamie: @Wizard can you find something that is specifically for service learning or volunteering?}	 
\item[] \texttt{[11:18] Wizard: I'm not finding a lot of service learning opportunities, especially in West Africa. I have a few from other regions, would you like to look at those?}	 
\end{itemize}
}

\emph{Wizard's comments:} \emph{``I wanted to tell them about the limitations of what I was finding and then offer a new strategy.''}

\subsubsection{\textbf{Checking-In [T]}}

The Wizard was able to monitor the participants' conversation in both the \textsc{BotDialog} and \textsc{BotTask} conditions. However, in the \textsc{BotTask} condition, the Wizard could also intervene to elicit feedback from the participants regarding their assistance. This happened when the Wizard wanted to: (1) check whether search results sent to participants were helpful and (2) check whether participants needed further assistance. Consider the following example from S01 (\textsc{BotTask}):

{\footnotesize 
\begin{itemize}[leftmargin=.05\linewidth,rightmargin=.05\linewidth]
\itemsep 0em 
\item[] \texttt{[11:22] Wizard: [Search Result] Do any of those recipes look good?}
\item[] \texttt{[11:50] Jamie: @Wizard the tomato and eggplant cooked salad does}
\item[] \texttt{[12:22 - 13:31] [Jamie and Alex discuss the difficulty of the recipe]}
\item[] \texttt{[13:50] Wizard: Ok. The directions also say that it will take 60 minutes to cook - does that fit in your schedule?}		
\end{itemize}
}

\emph{Wizard's comments:} \emph{``[I was] following up to make sure that we had fulfilled their goal of finding a recipe. I didn't want to leave them with something that they're only semi satisfied with.''}

\subsection{Theme: Repair Unsuccessful Interactions}

In some cases, the Wizard took initiative to repair previous interactions that were not successful. This happened when: (1) the participants misinterpreted the Wizard's question/suggestion or (2) the Wizard failed to communicate their intention through the intervention. In either case, the Wizard took initiative to ask another clarification question or rephrase a question/suggestion. 

\subsubsection{\textbf{Unexpected Response [D\&T]}}

Sometimes, participants misinterpreted the Wizard's question/suggestion and responded in a way that seemed \emph{unexpected} to the Wizard. For example, in S01 (\textsc{BotDialog}), the Wizard asked the participants about their \emph{purpose} for traveling (e.g., business, vacation), to which they responded with ``winter break''. While the response was technically correct, it did not match the intent of the Wizard's question:

{\footnotesize 
\begin{itemize}[leftmargin=.05\linewidth,rightmargin=.05\linewidth]
\itemsep 0em 
\item[] \texttt{[01:58] Wizard: what is the trip for?}
\item[] \texttt{[02:38] Alex: @Wizard winter break}	 
\item[] \texttt{[02:39] Wizard: for a vacation or is there another purpose?}	 
\end{itemize}
}

\emph{Wizard's comments:} \emph{``It seemed like they didn't really answer the question in the way that I was looking for... I didn't know if they were looking for a fun vacation or some other purpose."}

\subsubsection{\textbf{Sub-optimal Interventions [D\&T]}} 
In both the \textsc{BotDialog} and the \textsc{BotTask} conditions, the Wizard had to make dynamic decisions about their follow-up questions and suggestions (e.g., what to ask, how to ask) in real time. This dynamic nature of their work sometimes led them to make sub-optimal decisions (e.g., asking potentially ambiguous questions). In such situations, the Wizard tried to repair the failed intervention. In the \textsc{BotDialog} condition, they did this by asking an additional follow-up question. In the \textsc{BotTask} condition, they did this by clarifying their intention proactively. For example, in S15 (\textsc{BotTask}), the Wizard wrongly framed a question and attempted to fix the mistake upon realization: 

{\footnotesize 
\begin{itemize}[leftmargin=.05\linewidth,rightmargin=.05\linewidth]
\itemsep 0em 
\item[] \texttt{[03:48] Wizard: Do you want me to look for visa requirement for citizens of a certain country?}
\item[] \texttt{[04:37] Jamie: @Wizard Yes, can you look for visa requirements of brazil?}
\item[] \texttt{[04:50] Wizard: No, my question is: visas for people from what country?}
\end{itemize}
}

\emph{Wizard's comments:} \emph{``I was trying to clarify my previous question, because my previous question had been worded weirdly, which I sort of knew... I could tell that they misunderstood what I meant.''}

\section{Results --- RQ2: Timing Rationales}\label{sec:motivations_timing}

In this section, we present (sub)themes that characterize why the Wizard thought the \emph{timing} of an intervention was appropriate. These (sub)themes were derived from the Wizard's responses to the question: ``Why did you think the timing of this intervention was appropriate?''

\subsection{Theme: Urgency for Action}\label{subsec:time_urgency}

The Wizard justified the timing of an intervention based on a sense of urgency. In these cases, the benefits for intervening outweighed the risk of being perceived by participants as distracting or disruptive. In other words, the Wizard intervened with no delay to pursue an objective that was considered highly important at the time.

\subsubsection{\textbf{Clarify Immediately After a Request [D\&T]}} 

In some cases, the Wizard was unclear about a search request issued by participants. In such cases, the Wizard decided to ask a follow-up question to clarify the participants' intent before conducting a search rather than trying to resolve any uncertainties by themselves. For example, in S06 (\textsc{BotTask}), the participants issued a search request ``best travel destinations in South America'' and the Wizard immediately felt that the intent behind the phrase ``travel destinations'' needed clarification:  

{\footnotesize  
	\begin{itemize}[leftmargin=.05\linewidth,rightmargin=.05\linewidth]
		\itemsep 0em 
		\item[] \texttt{[02:47] Jamie: @Wizard best travel destinations in South America for Americans}
		\item[] \texttt{[03:08] Wizard:  By "travel destinations" do you mean countries or tourist attractions?}
	\end{itemize}
}

\emph{Wizard's comments:} \emph{``It's better than wasting my time trying to search for something that I know won't be accurate enough...if you're unsure about something, ask to clarify immediately after they ask because that might affect [the search].''}



\subsubsection{\textbf{Urgency to Provide an Update [D\&T]}}

In some cases, given a search request, the Wizard was unable to find relevant information within a reasonable time frame. In such cases, the Wizard experienced urgency to provide a status update. The tipping point for when the Wizard felt they had spent ``too much time'' on a search request was subjective and dependent on: (1) the nature of the search request (e.g., complexity) and (2) the participants' ongoing collaboration (e.g., moving forward with the task vs.~waiting idly for the Wizard's input). The Wizard felt that failing to provide an update was potentially more harmful than disrupting the participants' conversation. For example, in S17 (\textsc{BotTask}), the Wizard felt that the participants might be growing impatient because none of the previously sent results were helpful and they had been waiting long for the Wizard to respond to the latest search request.  Therefore, the Wizard wanted to give them ``something to look at'' while they continued searching.
 
{\footnotesize  
	\begin{itemize}[leftmargin=.05\linewidth,rightmargin=.05\linewidth]
		\itemsep 0em 
		\item[] \texttt{[02:38] Alex: @Wizard we're lokoing for a study abroad service learning opportunity in africa for unc undergrad students where we can partner with the community}
		\item[] \texttt{[03:01] Wizard: Are there countries or regions you're particularly interested in?}
		\item[] \texttt{[03:23] Jamie: @Wizard anything on the west africa side}
		\item[] \texttt{[04:56] Wizard: [Search Result]}
		\item[] \texttt{[04:57] Jamie: work concentrations, medical work concentrations...}
		\item[] \texttt{[05:36] Alex: aye. expensive}
		\item[] \texttt{[05:47] Jamie: not a fan. literally says traveling}
		\item[] \texttt{[07:13] Alex: we need to put service learning in there}
		\item[] \texttt{[07:25] Wizard: [Search Result] We could do a general search for service learning if you'd like!}
		\item[] \texttt{[08:52] Jamie: yeah that one didnt have much info for service learning or volunteering. maybe the research initiatives on the website}
		\item[] \texttt{[10:04] Alex: yeah i know, i just skipped over those. they didn't have much}
		\item[] \texttt{[11:01] Jamie: @Wizard a program that is low cost? in addition to the service learning}
		\item[] \texttt{[11:18] Wizard: I'm not finding a lot of service learning opportunities, especially in West Africa. I have a few from other regions, would you like to look at those? [Search Result]}
	\end{itemize}
}

\emph{Wizard's comments:} \emph{``This was where I thought less about that aspect of appropriate timing. Just because I've been working on it for a bit and I wanted to give them something to look at. It was just time to do that I guess because they were probably waiting on a source.''}

\subsubsection{\textbf{Urgency to Share Learned Information [T]}}

In the \textsc{BotTask} condition, the Wizard experienced urgency to intervene when they learned \emph{new} information that could affect the participants' approach to the task. For example, in S02, one participant suggested that December is the busiest time to visit Argentina. However, while searching, the Wizard learned that the busiest months to visit Argentina do not include December. Thus, they intervened to share this important information, as it might change how the participants thought about their options: 

{\footnotesize  
	\begin{itemize}[leftmargin=.05\linewidth,rightmargin=.05\linewidth]
		\itemsep 0em 
		\item[] \texttt{[12:46] Jamie: I mean christmas time is high traffic time. That's when people would be traveling to visit their international family}
		\item[] \texttt{[13:59] Wizard: The busiest months for tourism in Argentina are January, November, and May.}
		\item[] \texttt{[14:51] Jamie: @Wizard Check flight prices from the 14th to the 28th of december to buenos aires}
	\end{itemize}
}

\emph{Wizard's comments:} \emph{``It seemed like it was a good time to just get that information there so they would know it before we kept moving... so they have more information to consider.''} 

\subsubsection{\textbf{Urgency to Affirm [T]}} 

In the \textsc{BotTask} condition, the Wizard sometimes proactively intervened to affirm participants about a specific search strategy or decision. For example, in S11, when the Wizard noticed that the participants were finally considering a viable option (after struggling), they immediately intervened to encourage the participants to continue moving in this direction: 

{\footnotesize  
	\begin{itemize}[leftmargin=.05\linewidth,rightmargin=.05\linewidth]
		\itemsep 0em 
		\item[] \texttt{[10:26] Jamie: let's do 1-2 things in each city}
		\item[] \texttt{[13:05] Alex: we could do a bike tour in Sao Paulo}
		\item[] \texttt{[14:07] Wizard: Thinking about a bike tour is a good idea - you'll get more specific results if you know more what you're looking for.}
	\end{itemize}
}

\emph{Wizard's comments:} \emph{``...honing in on a particular type of activity [was] a good strategy in terms of searching and planning, so I wanted to affirm that is a good strategy. I was reacting to the conversation.''}

\subsection{Theme: Reacting to the Current Needs}

The Wizard also intervened when they noticed a change in the participants' needs and felt a need to react to it. This happened in two scenarios: (1) when participants started exploring new preferences and (2) when participants seemed to be struggling with the task. Importantly, in contrast to the subthemes in Section~\ref{subsec:time_urgency}, the following interventions were not motivated by a sense of urgency. Instead, they were motivated by the Wizard's desire to be seen as a valuable, attentive collaborator. Essentially, the Wizard viewed these interventions as an opportunity to be helpful and was less concerned with the intervention being distracting or disruptive.

\subsubsection{\textbf{Adjust Preferences [D\&T]}} 

In some cases, the Wizard noticed from the conversation that participants introduced new preferences that could impact the current search request. In such cases, the Wizard felt it was necessary to intervene to ask for clarification or provide a suggestion. The Wizard justified the timing of these interventions by wanting to be a valuable collaborator, capable of capturing the changing goals of the collaboration and redirecting their efforts accordingly. For example, in S25 (\textsc{BotTask}), participants first requested information about activities around the city of Buenos Aires. However, after the search request, participants expressed interest in the city of Mendoza and hiking. To react and adapt to the most current needs, the Wizard asked whether these new preference should be incorporated into the ongoing search:

{\footnotesize  
\begin{itemize}[leftmargin=.05\linewidth,rightmargin=.05\linewidth]
		\item[] \texttt{[13:34] Jamie: @Wizard how much are air bnbs in buenos aires?}
		\item[] \texttt{[14:18] Wizard: [Search Result]}
		\item[] \texttt{[14:41 - 16:43] [Jamie and Alex express their interest in going to other places around buenos aires such as the city of Mendoza and also mention hiking]}
		\item[] \texttt{[16:48] Wizard: Should we look for things about Mendoza and about hiking?}
\end{itemize}
}

\emph{Wizard's comments:} \emph{``They had sent a search query then they were having a conversation. I wanted to react to the things that they were talking about in that conversation...and incorporate and respond to those ideas when they were talking.''}

\subsubsection{\textbf{Resolve Challenges [T]}} 
In the \textsc{BotTask} condition, the Wizard sometimes intervened when they noticed participants were struggling with the task. In such cases, the Wizard felt they should promptly intervene to ask a follow-up question or offer a suggestion. For example, in S10, the Wizard noticed that participants were stuck on a specific option and felt that it was the right time to intervene to encourage more exploration: 

{\footnotesize  
	\begin{itemize}[leftmargin=.05\linewidth,rightmargin=.05\linewidth]
		\itemsep 0em 
		\item[] \texttt{[03:57] Jamie: or should we try Vietnamese pancake?
}
		\item[] \texttt{[03:59] Alex: should just make the food is shes hosting the party smh
}
\item[] \texttt{[04:40] Jamie: @Wizard What is in a Vietnamese pancake (Ban Xeo)?
}
		\item[] \texttt{[04:46] Wizard: Would you like some different recipe options to look at? you could make a decision about which kind of food after that.}
	\end{itemize}
}

\emph{Wizard's comments:} \emph{``The opportunity to widen the context and let them know there are other options out there because we were at a point where seeing different options would be helpful.''}

\subsection{Theme: Proximity as Appropriateness}
In some cases, the Wizard justified the timing of an intervention based on ``proximity'': (1) the intervention was close to the start of the task (during the initial exploration phase) and (2) the intervention was close to a search request issued by participants. Within these two ``proximity zones'', the Wizard felt they had permission to intervene without worrying about being disruptive. 

\subsubsection{\textbf{Proximity to Initial Exploration [T]}} 

The Wizard perceived their interventions to be less disruptive during the initial stages of exploration (i.e., before participants had committed to a particular approach to the task). To illustrate, consider the Wizard's comment from S25:

{\footnotesize  
\begin{itemize}[leftmargin=.05\linewidth,rightmargin=.05\linewidth]
		\item[] \texttt{[04:24] Jamie: @Wizard what is there for tourists to do in argentina?}
		\item[] \texttt{[04:44] Wizard: What kinds of activities would you want to do?}
\end{itemize}
}

\emph{Wizard's comments:} \emph{``The conversation just started [with a broad question]... So this conversation hadn't really gone in any other directions where it would be annoying [to intervene].''}

\subsubsection{\textbf{Proximity to the Search Request [D\&T]}}
The Wizard felt it was appropriate to intervene shortly after a search request. We observed this happening for two reasons. First, the Wizard thought they had the participants' joint attention, meaning that participants were not deeply engaged in other activities. Second, the Wizard perceived a search request as being equivalent to the participants handing them a ``talking stick''. In other words, the Wizard felt it was their turn to contribute to the collaboration. To illustrate, consider the Wizard's comment from S24 (\textsc{BotDialog}): 

{\footnotesize  
\begin{itemize}[leftmargin=.05\linewidth,rightmargin=.05\linewidth]
	\item[] \texttt{[05:02] Jamie: @Wizard How safe is Karongwe?}
	\item[] \texttt{[05:44] Wizard: Can you elaborate on what you mean by 'safe'?}
\end{itemize}
}

\emph{Wizard's comments:} \emph{``I thought it [the timing] was appropriate because they were in proximity of their original question. I thought it was close enough.''}

\subsection{Theme: Re-stimulate the Collaboration}

The Wizards also thought it was appropriate to intervene when they observed the participants' conversation slow down or divert into non-relevant topics. The Wizard interpreted both behaviors as a signal for them to get involved and re-stimulate the collaboration.

\subsubsection{\textbf{Peter Out [D\&T]}} \label{subsubsec:peter_out}

The Wizard thought it was appropriate to intervene when the pace of the conversation slowed down or became idle. The Wizard interpreted this ``petering out'' of the conversation as a sign of disengagement. For example, in S01 (\textsc{BotTask}), the Wizard noticed the participants' conversation slowed down. Thus, they did not think it would be disruptive to bring the participants' attention back to the information they shared earlier: 

{\footnotesize  
	\begin{itemize}[leftmargin=.05\linewidth,rightmargin=.05\linewidth]
		\itemsep 0em 
		\item[] \texttt{[13:13] Jamie: ooh the green bean saute}
		\item[] \texttt{[13:16] Alex: its all easily accessible}
		\item[] \texttt{[13:31] Jamie: i think we should do that it looks super simple}
		\item[] \texttt{[13:50] Wizard: Ok. The directions also say that it will take 60 minutes to cook - does that fit in your schedule?}
	\end{itemize}
}

\emph{Wizard's comments:} \emph{``So I think the conversation kind of petered out, and it was just a good time to jump in and make sure that was the recipe they wanted to look at and not something else.''} 

\subsubsection{\textbf{Side Conversation [D\&T]}} 
In some cases, the participants had conversations about something that was not relevant to the task. The Wizard interpreted these side conversations as a cue that an intervention would not be distracting or disruptive. To illustrate, consider the Wizard's comment from S22 (\textsc{BotTask}): 
{\footnotesize  
	\begin{itemize}[leftmargin=.05\linewidth,rightmargin=.05\linewidth]
		\itemsep 0em 
		\item[] \texttt{[03:33] Alex: yes animals would be cool}
		\item[] \texttt{[03:40] Jamie: okay cool}
		\item[] \texttt{[03:41] Alex: but voluntourism is not cool}
		\item[] \texttt{[03:55] Jamie: it is not. perhaps conservation effort then}
		\item[] \texttt{[03:56 - 04:12] [Alex asks Jamie if he has a preference about which country in West Africa they visit]}
		\item[] \texttt{[04:13] Wizard: I didn't find volunteer opportunities specifically for young adults, but I did find general volunteer opportunities in Africa. Would you like to see that? }
	\end{itemize}
}

\emph{Wizard's comments:} \emph{``They're talking about other ideas, just filling the time while waiting for me. They're doing a small talk. I figured they were going to bring their attention back to me whenever I did give them something. So I felt like it was okay to drop it in there.''}

\section{Discussion and Implications}\label{sec:discussion}

In this section, we summarize our findings and discuss their implications for designing mixed-initiative conversational search systems to support collaborators.  Our findings provides insights into what systems should consider when deciding whether and when to take initiative.

\textbf{Understanding of problem space:} The Wizard was motivated to intervene when they did not fully understand the collaborators' problem space (i.e., goals and constraints). This behavior was chiefly observed at the beginning of the task, when the Wizard only had an initial search request to make inferences about the problem space. Thus, when the search request provided insufficient \emph{contextual} information, the Wizard took the initiative to elicit more information about participants' goals and constraints. In such cases, the Wizard prioritized establishing a mutual understanding of the problem space with the participants over being potentially disruptive. 
An important question is: What makes a dynamic intervention disruptive?  Many studies have investigated this question from the perspective of a single person engaged in a task (see Li et al.~\cite{Li2012} for a review).  However, few studies have explored this question in the context of collaborative search~\cite{Avula2021}. Our findings suggest that at the beginning of the task, the need to establish common ground on task goals and constraints outweighs the risk of being disruptive. Therefore, systems should intervene during the early stages of a task if they lack information about the collaborators' problem space.  To realize this recommendation, future systems must be capable of: (1) inferring the current stage of a task, (2) detecting mentions of collaborators' goals and constraints in the conversation, and (3) intervening when important problem space information is unknown and undecided. 
 
\textbf{Proximity as heuristic:} Research has shown that there are phases during a task that are more amenable to interventions, such as during the beginning of the task \citep{Miyata} and transitions between subtasks \citep{Yoshiro}. These findings suggest that intervention timing can be justified \emph{in relation to} certain cues. In our study, the Wizard used the \emph{proximity} to a search request as a cue to determine if it was an appropriate time to intervene without being disruptive. The Wizard did not think twice about intervening shortly after a search request. We see two possible reasons for this trend: (1) the Wizard thought they had the participants' focus of attention and (2) the Wizard felt it was their turn to contribute to the collaboration. Additionally, the Wizard felt a sense of urgency to provide a status update if they took too long to respond to a search request. A natural question is: How long is ``too long''?  Based on the Wizard's comments, there is no predetermined threshold.  Instead, it is likely to depend on factors such as the complexity of the delegated task (e.g., the search request) and whether the collaborators are concurrently engaged in other aspects of the task versus being idle based on the conversation. Future work is needed to explore specific ways for systems to replicate both behaviors---intervening shortly after a request and providing a status update within a ``reasonable'' time period.  Providing a status update might also involve providing a suboptimal result while the system continues to search for a better result.

\textbf{Leveraging expectations:} The Wizard was motivated to intervene when they noticed a \emph{gap} between the participants' collaboration and their expectations of how a successful collaboration should proceed.  For example, consider some of the subthemes under ``Strategic Support''.  The Wizard intervened when participants: (1) had trouble decomposing the task into smaller, more manageable subtasks; (2) made search requests that seemed inconsistent with prior decisions; (3) had trouble focusing and narrowing the options to consider; and (4) lacked prerequisite knowledge needed to define the option space. In such cases, the Wizard took initiative to align the collaboration closer to their expectations of what an effective collaboration looks like.  Interestingly, in some cases, the Wizard asked clarification questions to not only better understand the participants' needs but \emph{also} to improve the quality of the collaboration. Future systems will need to have standards and criteria in order to detect issues in a collaboration.

\textbf{Leveraging the search results:} When intervening, the Wizard was often influenced by knowledge gained from search results they interacted with. For example, based on the search results, the Wizard recognized issues with a search request having: (1) ambiguous or underspecified intent; (2) subjective qualifiers (e.g. ``safe''); and (3) overly specific constraints that could be potentially dropped or modified. In response to these issues, the Wizard did a cost-benefit analysis to ask follow-up questions and/or provide suggestions. For instance, when the Wizard came across sub-optimal search results, they asked follow-up questions to improve the search results. However, if multiple follow-up questions did not improve the search results (i.e., diminishing returns), they shared a sub-optimal result and provided additional clarification. Also, the Wizard sometimes learned new things about the task topic as they searched.  For example, they learned about relevant alternatives that participants had not considered and they learned about important ``grouping criteria'' that could be used in a follow-up question to narrow the search space.  When the Wizard saw clear benefits from sharing this information, they explicitly made a suggestion or asked for clarification to \emph{nudge} the participants towards the information.

\textbf{Leveraging the conversation:} The Wizard leveraged their ability to monitor the participants' conversation.  We highlight four important ways in which the Wizard leveraged the conversation to decide whether and when to intervene.

First, the Wizard monitored the topic of the conversation.  They used this information in two ways.  First, they used it to ask follow-up questions and make suggestions that were \emph{on topic}.  Second, they used it to justify the point of intervention. The Wizard felt it was appropriate to intervene when participants talked about non-task-related topics. The Wizard took this as a sign that participants were not engaged with task-related activities and the intervention would not be disruptive.

Second, the Wizard captured and maintained an inventory of three specific things mentioned in the conversation: (1) options considered, (2) preferences stated, and (3) decisions made.  The Wizard used this information in five important ways.  First, the Wizard used this information to suggest options that participants had not considered but seemed relevant based on preferences stated in the conversation. Second, while processing a search request, the Wizard sometimes asked whether they should consider \emph{new} options/criteria mentioned in the conversation.  Third, the Wizard used the set of options being considered (and the lack of decisions made) to determine that participants were having trouble \emph{focusing}. Fourth, when participants started the task with only a narrow set of options, the Wizard justified their need to intervene to encourage participants to consider a wider range of options. Finally, the Wizard sometimes intervened when participants did something (e.g., issued a search request) that seemed \emph{inconsistent} with a previously made decision.  Importantly, this result suggests that future systems may be able to support collaborators by alerting them about actions that seem inconsistent with previously stated preferences and decisions made.

Third, the Wizard monitored \emph{how} participants communicated with each other. The Wizard used this information to provide support when participants seemed disengaged with the task or frustrated.

Finally, the Wizard monitored the \emph{pace} of the conversation.  The Wizard leveraged the pace of the conversation in several ways.  First, the Wizard noticed when the pace of the conversation slowed down and took this as a signal that participants needed support in re-engaging with the task.  Second, the Wizard noticed when participants spent too much time deliberating about options/criteria without making decisions.  The Wizard took this as a signal that participants needed help focusing.  Third, the Wizard felt compelled to provide a status update when they took too long to return information \emph{and} the participants seemed idle based on their conversation.

\textbf{Establish reliability:} Finally, some of the Wizard's interventions were motivated by them wanting to be perceived as a valuable and reliable collaborator. For example, the Wizard took initiative to: (1) explain how a search result could be used to support the participants' goals, (2) convey challenges encountered during a search, and (3) repair any unsuccessful interactions.  Interestingly, with respect to unsuccessful interactions, the Wizard was motivated to intervene regardless of whether the miscommunication was their fault or the participants'. Prior work in Human-AI collaboration also suggests that systems need to effectively convey reasons for their actions, as well as communicate their capabilities and limitations ~\cite{amershiHumanAi2019}. To realize such feedback mechanisms during collaborative search, systems must be capable of: (1) explaining their decisions, (2) detecting miscommunications, and (3) resolving miscommunications.

\textbf{Providing Socio-Emotional Support:} Prior work on agent-supported collaborations~\cite{kumar2011conversational,bales1950interaction,ai2010exploring,kumar2010socially} has studied the effects of two different types of agent interventions: instrumental (i.e., task-related) and expressive (i.e., socio-emotional).  Task-related interventions aim to help collaborators with specific aspects of the task.  On the other hand, expressive interventions aim to show solidarity, release tension, and motivate the group.  In our study, the Wizard also intervened with these two goals in mind.  In addition to providing task-related advice, the Wizard intervened to provide socio-emotional support.  For example, the Wizard intervened to ``check-in'' with the participants and ask whether they needed anything.  This was not done to directly influence the task, but rather to make the participants feel more comfortable with asking for assistance.  Additionally, the Wizard sometimes intervened when the pace of the collaboration was slow and they wanted to \emph{stimulate} the collaboration.  As described in detail below, in a previous paper~\cite{Avula2021} we reported on participants' perceptions of the Wizard during this same study. Participants reported that they appreciated the Wizard ``checking-in'' and that the Wizard helped keep the ``momentum'' of the collaboration. Thus, our results suggest that systems should sometimes proactively intervene to provide socio-emotional support and that such interventions may lead to positive perceptions.

\section{Exploring the Relations between The Wizard's Actions and the Participants' Perceptions} 

As mentioned in Section~\ref{sec:intro}, in a previous paper~\cite{Avula2021} we reported on results from the same study.  However, in that paper we focused exclusively on participants' perceptions of the Wizard and the task.  In this section, we try to make connections between the Wizard's actions (i.e., motivations and timing rationales) and participants' perceptions.  By joining both perspectives, we aim to better understand how an agent's goals might be perceived.  For example, are certain types of interventions more likely to be perceived as positive, negative, or either (depending on how the intervention is performed)?  

In terms of participants' perceptions, our results found that participants clearly perceived the \textsc{BotTask} condition differently than the \textsc{BotDialog} condition.  During the \textsc{BotTask} condition, participants reported lower levels of collaborative awareness and higher levels of disruption, frustration, and workload.  We believe that this resulted from the Wizard's \emph{proactivity} and their \emph{task-level suggestions}. Our results clearly show that the Wizard embraced their role of proactive collaborator.  They attempted to guide the collaboration by proactively making suggestions, sharing information, clarifying misconceptions, nudging participants to re-engage with the task if they lost focus, and even encouraging participants to pursue their current direction.  Having a third person in the loop might have caused participants to pay less attention to each other, resulting in lower collaborative awareness.  Similarly, at times, proactive task-level suggestions might have been perceived as disruptive and frustrating, especially if participants expected the Wizard to simply \emph{assist} and not \emph{guide} the collaboration. It might have not been a problem if participants had \emph{voluntarily} invited the Wizard to be a third group member with similar ``ground rules''. Prior work suggests that establishing ground rules is crucial in any collaborative setting~\cite{benotti2021,hilppo2010,lai2011} and the same appears to be true in a collaborative search setting. Future systems will need to learn how to clarify and negotiate roles and expectations with users.

\begin{figure}[h]
    \centering
    \includegraphics[width=\columnwidth]{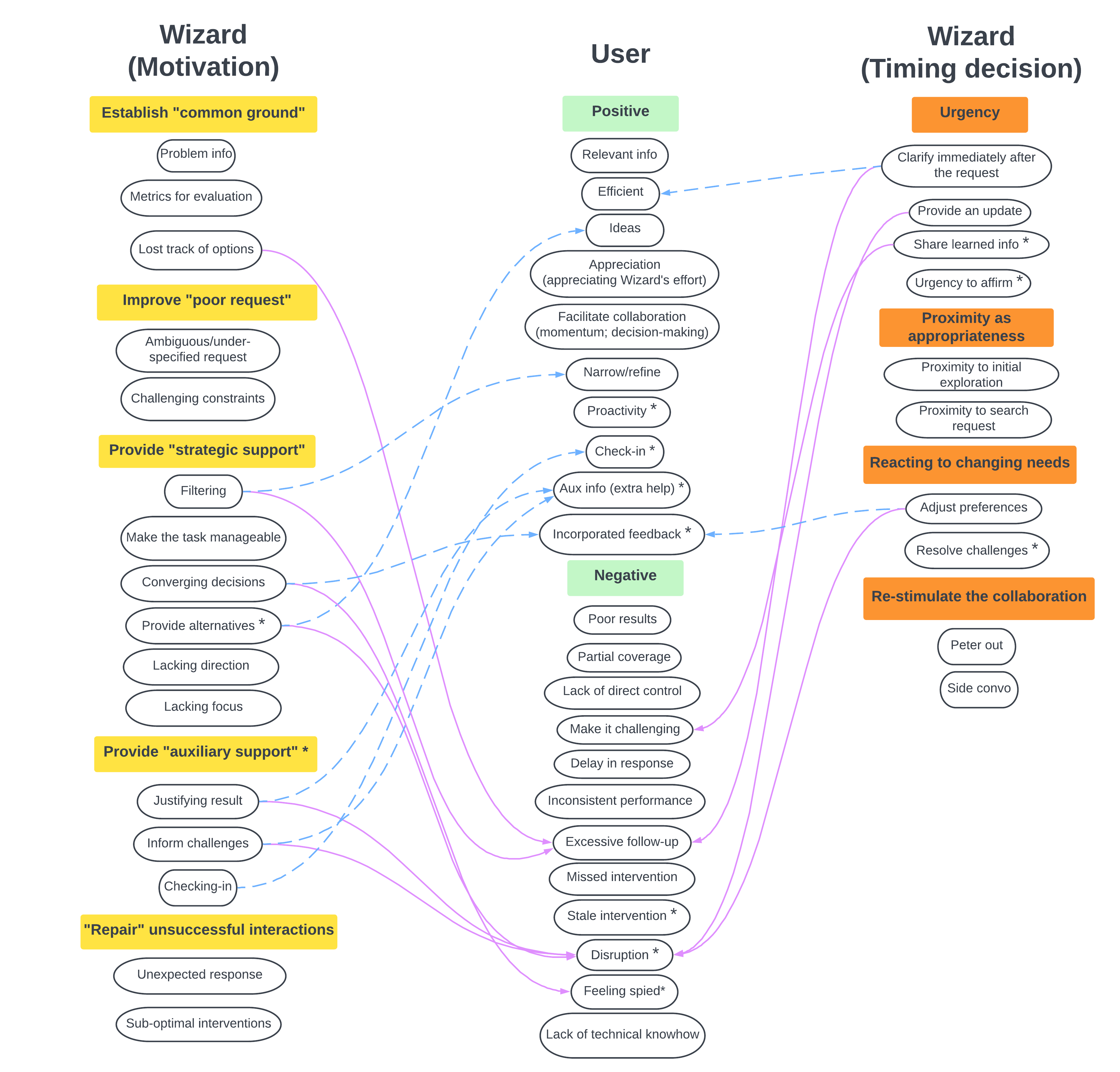}
    \caption{Connections between present and previous findings. Symbol `*' indicates that the theme was only observed in the \textsc{BotTask} condition. Dashed lines suggest positive implications and solid lines suggest negative implications of the Wizard's motivation and timing of intervention.}\label{fig:map}
\end{figure}

In our previous paper~\cite{Avula2021}, we reported on a \emph{qualitative} analysis of participants' responses to two open-ended questions about the Wizard---one about positive perceptions and one about negative perceptions.  In Figure~\ref{fig:map}, we draw connections between (sub)themes related to the Wizard's actions (i.e., motivations and timing rationales) and (sub)themes related to participants' perceptions of the Wizard.  It is important to note that these connections are \emph{speculative} albeit plausible.  Participants' open-ended responses did not enable us to connect every positive and negative perception with an action taken by the Wizard with 100\% certainty.  However, we discuss these \emph{plausible} connections to encourage future researchers to explore them in future work.  Importantly, some (sub)themes from the Wizard's perspective connect to negative (sub)themes from the participants' perspective.  These represent well-intended actions performed by the Wizard that may have resulted in negative perceptions.  Future research is needed to determine whether these actions either: (1) always lead to negative perceptions or (2) only lead to negative perceptions if they are done excessively, unnecessarily, or without proper justification.

In terms of positive perceptions, our results found that participants reported a wider \emph{range} of benefits from the Wizard in the \textsc{BotTask} versus \textsc{BotDialog} condition.  In fact, certain positive (sub)themes were only observed in the \textsc{BotTask} condition (marked with `*' in Figure~\ref{fig:map}).  Specifically, participants reported that they appreciated the Wizard being proactive, providing auxiliary information (i.e., extra help), and checking-in with participants in case they needed assistance.  These positive perceptions about the Wizard align with certain motivations and timing rationales reported in the present paper.  For example, the Wizard intervened in order to: (1) explain \emph{why} a specific search result was relevant and (2) explain challenges faced with a specific search request while suggesting an alternative direction to pursue.  These may have resulted in positive perceptions about the Wizard providing auxiliary information (i.e., extra help).

Our results also found that, despite the Wizard's good intentions, some interventions were perceived negatively.  For example, participants commented that the Wizard sometimes: (1) intervened excessively, (2) disrupted the collaboration, and (3) invaded their privacy. Our results in the present paper provide insights about actions taken by the Wizard that may have led to these negative perceptions.  

Some of the Wizard's interventions were motivated by \emph{long-term} goals. We use the term ``long-term'' to mean that the Wizard was sometimes more concerned about the quality of the end result than about satisfying the participants' immediate needs. For example, the Wizard often intervened to learn about the participants' \emph{problem space} (i.e., task-level goals, preferences, and constraints).  The Wizard elicited this information because it influences relevance criteria for the current search request and all subsequent search requests during the task.  However, the long-term benefits of these interventions may not have been fully understood and appreciated by participants.

Additionally, some of the Wizard's interventions were motivated by a sense of \emph{urgency}.  For example, the Wizard intervened when they felt it was necessary to provide an update as they worked on a search request or to correct an important misconception that participants had.  In such cases, the Wizard intervened because they thought it was important---the risk of being disruptive was secondary.  It is possible that participants did not fully appreciate the importance of such interventions.

Finally, during the task, the Wizard monitored the participants' conversation.  This enabled the Wizard to clarify misconceptions, suggest alternatives that had not been considered, and incorporate preferences mentioned in the conversation when processing a search request.  Despite these benefits, participants also reported feeling a loss of privacy.

Together, these results suggest that mixed-initiative systems should: (1) explain the motivation behind an intervention, (2) explain the timing of an intervention if it is likely to be disruptive, and (3) allow users to control the amount of monitoring being done. More generally, future research is needed to investigate the discrepancy between the intentions of an agent and the perceptions of users.

\section{Caveats \& Limitations}\label{sec:limitations}
Our study involved two experimental decisions that may have influenced our results.

\textbf{Task time limit:}  During the study, participants were given 15 minutes to complete each task.  This time limit was imposed to keep the study session under two hours. This time limit may have influenced the Wizard's motivations for taking initiative. For example, in the \textsc{BotTask} condition, the Wizard sometimes intervened with task-level advice when they felt that participants were not making sufficient progress on task.  Such interventions may have been influenced by the Wizard knowing that participants had 15 minutes to complete the task. In other words, the Wizard might have taken a more conservative approach if participants had not had a time limit.

\textbf{Search only with the help of the searchbot:}  During the study, participants could not search individually on their own browsers and had to gather information by sending search requests to the searchbot from Slack.  We made this decision to ensure a certain level of engagement between participants and the searchbot.  Naturally, in some real-world situations, collaborators may also be able to search on their own.  The impact of this decision is an open question for future work.  Future studies could extend our research and consider a set-up in which collaborators can interact with a conversational search system and also search independently on their own.

\textbf{Connections between the Wizard's motivations and the participants' perceptions:} In Section~\ref{sec:discussion}, we discuss \emph{plausible} connections between the Wizard's motivations/rationales and the participants' positive and negative perceptions of the Wizard (reported in Avula et al.~\cite{Avula2021}).  While we asked about the Wizard's motivations/rationales for every intervention, we asked about participants' perceptions of the Wizard at the session-level (i.e., in a post-task questionnaire).  For this reason, the connections reported in Section~\ref{sec:discussion} are speculative.  We discuss these plausible connections to encourage researchers to explore them more directly in future work. Research should investigate how well-intended decisions by a collaborative agent may lead to negative experiences by collaborators.

\section{Conclusion}\label{sec:conclusion}

In this paper, we explored how a conversational search agent can support people during a collaborative search task performed over an instant messaging platform (i.e., Slack).  We reported on a ``Wizard of Oz'' study in which a reference librarian played the role of a conversational agent that can take different levels of initiative: dialog- and task-level initiative.  We presented results from a qualitative analysis of stimulated recall interviews performed with the Wizard. Our analysis focused on understanding the Wizard's motivations for taking initiative and their rationales for the timing of each intervention. Our findings provide insights into why and when a mixed-initiative agent should intervene during a collaborative search task.  Finally, we discussed implications for designing conversational search systems to support collaboration.  Our work extends prior studies by bridging two important areas of ongoing research: conversational and collaborative search. Our objective was to gain insights about the ``action space'' of an agent operating at the intersection of these two fields.

\begin{acks}
This work was supported in part by Edward G. Holley research grant and NSF grant IIS-1451668. Any opinions, findings, conclusions, and recommendations expressed in this paper are the authors' and do not necessarily reflect those of the sponsors.
\end{acks}

\bibliographystyle{ACM-Reference-Format}
\bibliography{system_initiative_cscw}

\end{document}